\documentclass[granma]{svjour}
\usepackage{graphics,epsfig,citesort,psfrag,amssymb}

\newcommand{\vn}[1]{\mbox{\boldmath$#1$}}
\newcommand{\bsigma}{\vn{\sigma}}
\newcommand{\bvarepsilon}{\vn{\epsilon}}
\newcommand{\bepsilon}{\vn{\varepsilon}}
\newcommand{\tr}{{\rm tr}}
\newcommand{\dev}{{\rm dev}}
\newcommand{\grad}{\rm grad\,}
\renewcommand{\div}{\rm div\,}
\renewcommand{\matrix}[1]{{\bf{#1}}}

\begin{document}

\title{Macroscopic material properties from quasi-static, microscopic
  simulations of a two-dimensional shear-cell}

\author{Marc L\"atzel, Stefan Luding, and Hans J. Herrmann (*)}
\institute{          
           Institute for Computer Applications 1, \\
           Pfaffenwaldring 27, \\
           70569 Stuttgart, GERMANY\\ 
           Correspondence to: m.laetzel@ica1.uni-stuttgart.de\\
           (*) PMMH, ESPCI, 10 rue Vauquelin, 75231 Paris, France\\
           ~\\
\noindent
We thank S. Diebels, W. Ehlers, D. Schaeffer, J. Socolar, O. Tsoungui 
and W. Volk for helpful and inspiring discussions. Also we appreciate
the cooperation with B. Behringer and D. Howell and acknowledge the support
of the the German National Science Foundation DFG through the research
group `Modellierung koh\"asiver Reibungsmaterialien'.  }
%{\em Received: 23 July 1999}
\maketitle

{\bf \noindent Abstract~} 
One of the essential questions in the area of granular matter is, how
to obtain macroscopic tensorial quantities like stress and strain from
``microscopic'' quantities like the contact forces in a granular 
assembly. Different averaging strategies are introduced, tested, 
and used to obtain volume fractions, coordination numbers, and 
fabric properties. We derive anew the non-trivial relation for 
the stress tensor that allows a straightforward calculation of the 
mean stress from discrete element simulations and comment on the 
applicability.  
Furthermore, we derive the ``elastic'' (reversible) mean displacement 
gradient, based on a best-fit hypothesis. Finally, different combinations 
of the tensorial quantities are used to compute some material properties. 

The bulk modulus, i.e.~the stiffness of the granulate, is a linear
function of the trace of the fabric tensor which itself is
proportional to the density and the coordination number. The fabric,
the stress and the strain tensors are {\em not} co-linear so that a 
more refined analysis than a classical elasticity theory is 
required.

%\tableofcontents
\section{Introduction}

Macroscopic continuum equations for the description of the behavior
of granular media rely on constitutive equations for stress, strain,
and other physical quantities describing the state of the system.
One possible way of obtaining an observable like the stress is to perform
discrete particle simulations \cite{herrmann98,cundall79}
and to average over the ``microscopic'' quantities in 
the simulation, in order to obtain an averaged macroscopic
quantity. In the literature, slightly different
definitions for stress and strain averaging procedures can be found 
\cite{drescher72,rothenburg81,savage81,cundall82,goddard86,bathurst88,kruyt96,liao97b,goldhirsch99,kuhl99}. 

The outline of this study is as follows.  In section~\ref{sec:methods}
the discrete element simulation method is discussed and, in
section~\ref{sec:micro-macro} some averaging methods are introduced
and applied to a scalar quantity, namely the volume fraction.
Section~\ref{sec:macro} contains the definitions and averaging
strategies for fabric, stress, and elastic strain and in
section~\ref{sec:results} some material properties are extracted from
the results obtained in section~\ref{sec:macro}.

\section{Modelling discrete particles}
\label{sec:methods}

The elementary units of granular materials are mesoscopic grains. In
order to account for the excluded volume, one can assume that the
grains are impenetrable but deform under stress.  Since the realistic
modelling of the deformations of the particles in the framework of a
continuum theory \cite{walton93,lian96} would be much too complicated,
we relate the interaction force to the overlap $\delta$ of two
particles. %(see Fig.~\ref{fig:deform}). 
Note that the evaluation of the inter particle forces based on the
overlap may not be sufficient to account for the nonlinear stress
distribution inside the particles. Consequently, our results presented
below are of the same quality as this basic assumption. 

The force laws used are
material dependent, involving properties such as Young's modulus of
elasticity, and have to be validated by comparison with experimental
measurements \cite{foerster94,labous97,falcon98}.
%\begin{figure}[ht]
%\begin{center}
%\epsfig{figure=deform.eps,height=2cm}
%\end{center}
%\caption{Schematic drawing of the overlap $\delta$, 
%interpreted as an effective deformation of the particle contacts.
%The solid lines indicate the particle outlines in the overlapping
%situation, the dashed line the ``virtual'' deformation on contact.}
%\label{fig:deform}
%\end{figure}

When all forces $\vec f_i$ acting on the particle $i$, either from other
particles, from boundaries or from external forces, are known, the
problem is reduced to the integration of Newton's equations of motion
for the translational and the rotational degrees of freedom
\begin{equation}
m_i \frac{{\rm d}^2}{{\rm d} t^2} \vec r_i = \vec f_i~, {\rm ~~and~~}
I_i \frac{{\rm d}^2}{{\rm d} t^2} \vec \Phi_i = \vec M_i 
~.
\label{eq:newton_trans}
\end{equation}
The mass of particle $i$ with diameter $d_i$ is $m_i$, and its moment
of inertia is $I_i=q_i m_i (a_i)^2$, with the radius $a_i=d_i/2$ and
the dimensionless shape factor $q_i$. 
The vectors $\vec r_i$ and $\vec \Phi_i$ give the
position and the orientation angles of particle $i$, respectively. In our
model attractive forces and the presence of other phases are
neglected, we focus on ``dry granular media''.  Particle-particle 
interactions are short range and active on contact only. The total 
force (torque) due to other particles is thus 
$\vec f_i = \sum_c \vec f^c_i$ ($\vec M_i = \sum_c \vec M_i^c$), 
where the sum runs over all contacts of particle $i$.
The torque $\vec M_i^c = \vec l_i^c \times \vec f_i^c$ is related to
the force $\vec f_i^c$ via the cross product with the branch vector $\vec
l_i^c$ from the particle center to the point of contact $c$. Eq.~(\ref{eq:newton_trans}) consists of six scalar equations in three
dimensions and reduces to three equations in two dimensions (2D), two for the
linear and one for the rotational degree of freedom.  In the following
the force laws for $\vec{f}_i^c$ accounting for excluded volume,
dissipation, and friction are introduced.

\subsection{Force laws}
\label{sec:forcelaws}

The particles $i$ and $j$ interact only when they are in contact so that their
overlap $\delta = {1 \over 2}(d_i + d_j) - (\vec r_i - \vec r_j) \cdot
\vec n$ is positive, with the unit vector 
$\vec n = (\vec r_i - \vec r_j) / |\vec r_i - \vec
r_j|$ that points from $j$ to $i$. The symbol `$\cdot$' denotes
the scalar product of vectors or, more generally, the order-reduction 
by one for each of two tensors.

The first contribution to the force acting on particle $i$ from $j$ is
an elastic repulsive force
\begin{equation}
\vec f_{\rm n,el} = k_n \delta \vec n~,
\label{eq:fn}
\end{equation}
where $k_n$ is proportional to the material's modulus of elasticity
with units $[N/m]$. Since we are interested in disks rather than
spheres, we use a linear spring that follows Hooke's law,
whereas in the case of elastic spheres, the Hertz contact law would be
more appropriate \cite{hertz82,landau75}.

The second contribution, a viscous dissipation, is given by the 
damping force in the normal direction
\begin{equation}
\vec f_{\rm n,diss} = \gamma_n \dot \delta \vec n~,
\label{eq:fdiss}
\end{equation}
where $\gamma_n$ is a phenomenological normal viscous dissipation coefficient 
with units [kg s$^{-1}$] and $\dot \delta = - \vec v_{ij} \cdot \vec n$
the relative velocity in the normal direction 
$\vec v_{ij} = \vec v_i - \vec v_j$. 

The third contribution to the contact force -- accounting for
tangential friction -- can be chosen in the simplest case, according
to Coulomb, as $\vec{f}_{\rm t,friction} \le -\mu|\vec{f_n}|\vec{t}$,
where $\mu$ is the friction coefficient and $\vec t = \dot{\vec{\xi}}
/ |\dot{\vec{\xi}}|$ is the tangential unit-vector parallel to the
tangential component of the relative velocity
$\dot{\vec{\xi}}=\vec v_{ij}-(\vec v_{ij}\cdot\vec{n})\vec{n}$.
Because this non-smooth ansatz leads to numerical problems for small
$\dot{\vec{\xi}}$ a regularizing viscous force $\vec{f}_{\rm
  t,viscous}=-\gamma_t\dot{\vec{\xi}}$ is added. The two forces are
combined by taking the minimum value
\begin{equation}
  \vec{f}_{\rm t}=-{\rm
      min}(|\gamma_t\dot{\vec{\xi}}|,|\mu\vec{f_n}|)\vec{t} ~.
  \label{eq:friction_force}
\end{equation}
The effect of a more realistic tangential force law according to
Cundall and Strack \cite{cundall83} will be reported 
elsewhere \cite{latzel99a}. 
Due to the boundary conditions introduced below, it is also necessary
to account for the friction with the bottom
\begin{equation}
  \vec{f}_{\rm b}=- \mu_{\rm b}mg \hat {\vec v}~,
  \label{bottomfric}  
\end{equation}
with the unit vector in the direction of the particles velocity
$\hat {\vec v} = \vec{v}/|\vec{v}|$.  The effect of a bottom friction 
was discussed also in \cite{schollmann99}. In summary, we combine the 
forces and use
\begin{equation}
  \label{summforces}
  \vec{f}_{i} = \sum_c (\vec{f}_{\rm n,el}+\vec{f}_{\rm n,diss}+
                \vec{f}_{\rm t} ) +\vec{f}_{\rm b}
\end{equation}
for the forces acting on particle $i$ at its contact $c$ with particle
$j$.

\subsection{Model system}

\begin{figure}[htb]
%\psfrag{inner ring}{$R_i=0.1032$\,m}
%\psfrag{outer ring}[r][r]{$R_o=0.2524$\,m}
\psfrag{inner ring}{$R_i$}
\psfrag{outer ring}[r][r]{$R_o$}
\psfrag{25.24 cm}{\tiny 25.24 cm}
\psfrag{10.32 cm}{\tiny 10.32 cm}
\psfrag{disks}{}
\psfrag{2911}{}
\begin{center}
\epsfig{file=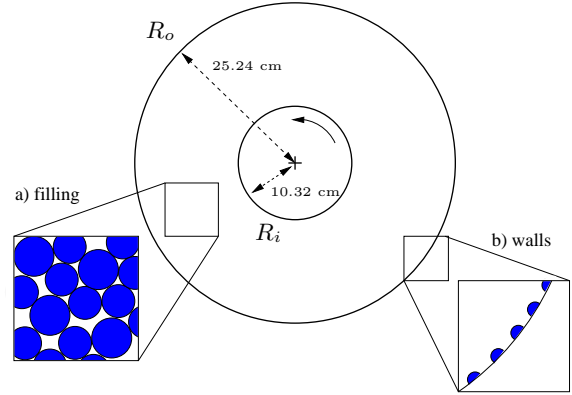,height=5.2cm,angle=0}
\end{center}
\caption{A schematic plot of the model system}
\label{fig:shearcell}
\end{figure}

In the simulations presented in this study, a two-dimensio\-nal Couette
shear-cell is used, filled with a bidisperse packing of disks, as sketched in
Fig.~\ref{fig:shearcell}. The system undergoes slow shearing
introduced by turning the inner ring. The properties of the particles,
used for the force laws of subsection~\ref{sec:forcelaws},
are summarized in table~\ref{tab:model_props}.
The boundary conditions are based on an experimental 
realization~\cite{howell99,veje98b,veje99}.
For more details on other simulations, see~\cite{veje98b,schollmann99}.
\begin{table}[htb]
\begin{center}
\begin{tabular}{|c||c|}
\hline
property                              & values\\
\hline
\hline
diameter $d_{\rm small}$, mass $m_{\rm small}$ 
                                      & $7.42$\,mm, $0.275$\,g\\
diameter $d_{\rm large}$, mass $m_{\rm large}$
                                      & $8.99$\,mm, $0.490$\,g\\
wall-particle diameter $d_{\rm wall}$,& $2.50$\,mm \\
system/disk-height $h$                & $6$\,mm \\
normal spring constant $k_n$          & $352.1$\,N/m \\
normal viscous coefficient $\gamma_n$ & $0.19$\,kg/s \\
tangential viscous damping $\gamma_t$ & $0.15$\,kg/s \\
Coulomb friction coefficient $\mu$    & $0.44$\\
bottom friction coefficient $\mu_b$   & $2\times 10^{-5}$ \\
material density $\rho_0$             & $1060$\,kg\,m$^{-3}$ \\
\hline
\end{tabular}
\end{center}
\caption{Microscopic material parameters of the model.}
\label{tab:model_props}
\end{table}

In the simulations different global volume fractions
\begin{equation}
  \label{eq:rho}
  \bar \nu=\frac{1}{V_{\rm tot}}\sum_{p=1}^N V^p
\end{equation}
of the shear-cell are examined. The sum in Eq.~(\ref{eq:rho}) runs 
over all particles $p$ with volume $V^p$ in the cell with 
$V_{\rm tot} = \pi (R_o^2-R_i^2)$.
In this study $\bar \nu$ is varied between $0.8084$ and $0.8194$, by
varying the particle number, see table~\ref{tab:sim_details}. For
the calculation of the global volume fraction, the small particles
glued to the wall are counted with half their volume only, and thus
contribute with $\bar \nu_{\rm wall} = 0.0047$ to $\bar \nu$.
\begin{table}[htb]
\begin{center}
\begin{tabular}{|c|c|c|c|c|c|}
\hline
 &  global volume       & \multicolumn{2}{c|}{number of particles} & $t_{\rm max}$ & $\Delta t$\\
 &  fraction $\bar \nu$ & \ \ small \ \ & large                    & $(s)$         & $(s)$    \\
\hline
\hline
A &  0.8084 & 2511  &   400 & 335 & 5\\
B &  0.8103 & 2520  &   399 & 119 & 1\\
C &  0.8133 & 2524  &   404 & 119 & 1\\
D &  0.8149 & 2545  &   394 & 119 & 1\\
E &  0.8180 & 2538  &   407 & 505 & 5\\
F &  0.8194 & 2555  &   399 & 119 & 1\\
\hline
\end{tabular}
\end{center}
\caption{Details of the simulation runs studied in this paper.}
\label{tab:sim_details}
\end{table}

Note that we use the phrases ``volume'' and ``volume fraction'' even if,
strictly speaking, the unfamiliar terms ``disk area'' and ``area fraction'' 
could be used. The reasons for this choice are:
(i) The methods discussed in this study are straightforwardly generalized
to three dimensions and (ii) the particles are three dimensional objects
with height $h$ anyway, so that the use of the word ``volume'' is justified. 

\subsection{Initial conditions and steady state}

The simulations are started in a dilute state with an extended outer
ring $R_o(t=0) > R_o=0.2524$\,m, and the inner ring already rotates 
counterclockwise with constant angular frequency $\Omega = 2 \pi / T_i 
= 0.1$\,s$^{-1}$ and period $T_i=62.83$\,s. The extended outer ring
is used in order to allow for a random, dilute initial configuration. 
The desired density is then reached by reducing the volume. The radius 
of the outer ring is reduced within about two seconds to reach its
desired value $R_o$. Afterwards, the outer ring is kept fixed and the inner 
ring continues to rotate until at $t=t_{\rm max}$ the simulation ends.
If not explicitly mentioned, averages are performed after about one 
rotation at $t=60$\,s (to get rid of the arbitrary initial configuration), 
and during about one rotation, until $t=119$\,s.

\section{From the micro- to a macro-description}
\label{sec:micro-macro}

In the previous section, the microscopic point of view was introduced,
as used for the discrete element method. Particles are viewed as
independent entities which interact when they come in contact. In this
framework, the know\-ledge of the forces at each contact is sufficient
to model the dynamics and the statics of the system. Tensorial
quantities like the stress or the deformation gradient are not
necessary for a discrete modelling. However, subject of current research
is to establish a correspondence to continuum theories by computing
tensorial quantities like the stress $\bsigma$, the strain $\bepsilon$, 
as well as scalar material properties like, e.g., the bulk and shear moduli
\cite{goddard86,kruyt96,liao97b}. In the course of this process, we
first discuss averaging strategies using the material density as an
example. 

\subsection{Averaging strategies}

Most of the measurable quantities in granular materials vary strongly on 
short distances. 
%As an example, the stress is not constant inside a grain
%but has its largest value at the contacts. 
Thus, computing averages necessitates dealing with or smearing out the
fluctuations. In order to suppress the fluctuations, we perform
averages in both time and space.  This is possible due to the chosen
boundary condition. The system can run for long time in a quasi-steady
state and, due to the cylindrical symmetry, points at a certain
distance $R$ from the origin are equivalent to each other. Therefore,
averages are taken over many snapshots in time with time steps $\Delta
t$ and on rings of material at a distance $r=R-R_i$ from the inner ring.
The width of the averaging rings is
$\Delta r$, so that the averaging volume of one ring is $V_r=2 \pi r
\Delta r$. For the sake of simplicity (and since the procedure is not
restricted to cylindrical symmetry), the averaging volume is denoted
by $V=V_r$ in the following. The rings are numbered from $s=0$ to
$B-1$, with $B=(R_o-R_i)/\Delta r$, and ring $s$ reaches from
$r_s=r-\Delta r/2$ to $r_{s+1}=r+\Delta r/2$.  The averaging over many
snapshots is somehow equivalent to an ensemble average.  However, we
remark that different snapshots are not necessarily independent of
each other as discussed in subsection~\ref{sec:time_av}.  Also the
duration of the simulation maybe not long enough to explore a
representative part of the phase space.

%Because the system is
%believed to be in quasi-static equilibrium we compute ``snapshots'' of
%the system for different times and thus average over many
%realizations. On the other hand we take advantage of the fact that our
%model system has a cylindrical symmetry. Thus we divide the system in
%equally spaced rings and use this rings as the reference volume
%introduced in the previous sections.
%\begin{figure}[ht]
%\begin{center}
%\epsfig{figure=new_results/binning_stress.eps,height=5.4cm}
%\end{center}
%\caption{Results for the stress in radial direction. For small
%  binnings the values converge.}
%\label{fig:binn_stress}
%\end{figure}
%\begin{figure}[ht]
%\begin{center}
%\epsfig{figure=new_results/binning_deviator.eps,height=5.4cm}
%\end{center}
%\caption{Results for the stress deviator for different binning sizes.}
%\label{fig:binn_deviator}
%\end{figure}

The cylindrical symmetry is accounted for by a rotation of all
directed quantities like vectors, depending on the cartesian position 
$\vec r_i = (x_i,y_i)$ of the corresponding particle $i$. The 
orientation of particle $i$ is $\phi_i = \arctan(y_i/x_i)$ for
$x_i>0$ and periodically continued for $x_i<0$ so that
$\phi_i$ can be found in the interval $[-\pi,\pi]$. The vector
$\vec n^c$ that corresponds to contact $c$ of particle $i$ is
then rotated about the angle $-\phi_i$ from its cartesian orientation
before being used for an average. Note that this does 
{\em not} correspond to a transformation into orthonormal 
cylindrical coordinates.
%In the following, the index $r$
%is used for the radial outward direction and the index $\phi$ is 
%used for the counterclockwise perpendicular direction. Note that 

Finally, we should remark that the most drastic assumption used for
our averaging procedure is the fact, that all quantities are smeared
out over one particle. Since it cannot be the goal to solve for the
stress field inside one particle, we assume that a measured quantity
is constant inside the particle. This is almost true for the density,
but not e.g.~for the stress.  However, since we average over all
positions with similar distance from the origin, i.e.~averages are
performed over particles with different positions inside a ring, details 
of the position dependency inside the particles will be smeared out anyway.
An alternative approach was recently proposed by I. Goldhirsch 
\cite{goldhirsch99} who smeared out the averaging quantities along the 
lines connecting the centers of the particles and weighed the contribution
according to the fraction of this line within the averaging volume.

\subsection{Volume fractions}

The first quantity to measure is the local volume fraction
\begin{equation}
  \label{eq:nureal}
  \nu = \nu(r) = \frac{1}{V}\sum_{p \in V} w_V^p V^p
\end{equation}
with the particle volume $V^p$. 
%In the following the brackets 
%$\langle Q \rangle$ denote the average over a quantity $Q$ of the particles 
%in an averaging volume $V$. Since only the particles contribute, but not the 
%pore-space, one has $\langle 1 \rangle = \nu$, i.e.~the volume fraction
%enters the averaged values. 
$w_V^p$ is the weight of the 
particle's contribution to the average. 
%At this point, one has 
%several possibilities to choose the weight $w_V^p$ that accounts for the 
%particle's contribution to the average.

This formalism can be extended to obtain the mean value of a quantity
$Q$ in the following way:
\begin{equation}
Q = \langle Q^p \rangle = \frac{1}{V} \sum_{p \in V} w_V^p V^p Q^p ~,
\end{equation}
with the pre-averaged particle quantity
\begin{equation}
Q^p = \sum_{c=1}^{{\cal C}^p} Q^c ~.
\end{equation}
with the quantity $Q^c$ attributed to contact $c$ of particle $p$.
In the following the brackets 
$\langle Q \rangle$ denote the average over a quantity $Q$ of the particles 
in an averaging volume $V$.

The simplest choice for $w_V^p$ is to use $w_V^p=1$,
if the center of the particle lies inside the ring, and $w_V^p=0$
otherwise.  This method will be referred to as {\em particle-center
averaging} in the following. 
%In the situation shown in Fig.\ 
%\ref{fig:av_vol}, particle-center averaging will account for all 
%particles sketched as thick circles.
%\begin{figure}[ht]
%\begin{center}
%\epsfig{file=rev1.eps,height=4.4cm,angle=0}
%\end{center}
%\caption{Schematic plot of discrete particles. The averaging volume
%is here the shaded circle and the particles plotted as thick circles
%contribute to the average.}
%\label{fig:av_vol}
%\end{figure}

\begin{figure}[ht]
\begin{center}
\epsfig{file=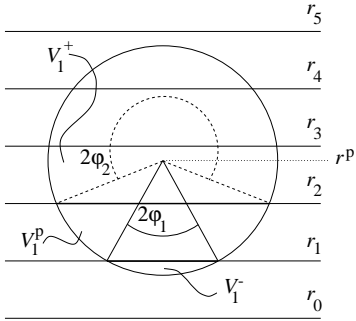,height=4.2cm,angle=0}
\end{center}
\caption{Schematic plot of a particle $p$ at radial position $r^p$
which is cut into pieces by the boundaries $r_s$ of the averaging volumes.
We assume $s=0$, $\ldots$, ${m+1}$ such that all $r_s$ with $s=1$, 
$\ldots$, $m$ hit the particle, i.e. $|r^p-r_s|<d/2$.
}
\label{fig:slices}
\end{figure}

A more complicated way to account for those parts of the particles 
which lie partially inside a ring is to use only the fraction of the
particle volume that is covered by the averaging volume. 
% , the grey area in 
%Fig.\ \ref{fig:av_vol}.  
Since an exact calculation of the area
of a disk that lies in an arbitrary ring is rather complicated, 
we assume that the boundaries of $V$ are locally straight, i.e.~we cut
the particle in slices, as shown in Fig.~\ref{fig:slices}. This method
will be referred to as {\em slicing} in the following. The error introduced 
by using straight cuts is well below one per-cent in all situations considered 
here.
 
The volume $V^p_s = w_V^p V^p$ of a particle $p$ which partially
lies between $r_s$ and $r_{s+1}$ is calculated by subtracting the 
external volumes $V_s^-$ and $V_s^+$  from the particle volume 
$V^p=\pi (d/2)^2$ so that
\begin{eqnarray}
V^p_s & = & V^p - V_s^- - V_s^+  \nonumber \\
      & = & (d/2)^2 [ \pi - \phi_s + \sin(\phi_s) \cos(\phi_s) \nonumber \\
      &   & ~~~~~~~~~~~   - \phi_{s+1} + \sin(\phi_{s+1}) \cos(\phi_{s+1}) ]
\end{eqnarray}
with $\phi_s = \arccos( 2(r^p-r_s)/d )$ 
 and $\phi_{s+1} = \arccos( 2(r_{s+1}-r^p)/d )$.
The term $(d/2)^2 \phi$ is the area of the segment of the circle
with angle $2 \phi$, and the term  $(d/2)^2 \sin(\phi) \cos(\phi)$
is the area of the triangle belonging to the segment.
In Fig.~\ref{fig:slices} the case $s=1$ is highlighted, and the boundaries 
between $V_s^-$, $V^p_s$, and $V_s^+$ are indicated as thick solid lines. 
The two outermost slices $V^p_0=V_1^-$ and $V^p_s=V_{s-1}^+$ have to be 
calculated separately. 

%In Fig.\ \ref{fig:dens1} the two averaging strategies 
When the two averaging strategies (particle-center and slicing)
are compared for different widths $\Delta r$ of the 
averaging rings, so that the number of intervals is $B=20$, 40, or 60. 
For $\Delta r \approx d_{\rm small}$ ($B=20$), the results
are almost independent of the averaging procedure. For finer binning 
$B_s\ge30$, the slicing procedure converges on a master curve with 
weak oscillations close to the walls which are due to the 
wall-induced layering of the particles. The results of the particle-center
averaging strongly fluctuate for $B_c\ge24$. These
oscillations come from ordered layers of the particles close to the walls
so that the slicing method reflects the real density distribution for
fine enough binning. The particle-center method, on the other hand, leads
to peaks, where the centers of the particles in a layer are situated and 
to much smaller densities where few particle centers are found; 
the particle-center density is obtained rather than the material density.
%\begin{figure}[ht]
%\begin{center}
%\epsfig{file=dens1.ps,height=7.5cm,angle=-90}
%\end{center}
%\caption{Volume fraction $\nu$ for different averaging methods
%and bin-widths $\Delta r=(R_o-R_i)/B_{c,s}$ from simulation D.
%The open symbols correspond to the particle-center criterion, the
%solid symbols to the slicing method with the respective number of
%rings $B_{c,s}$.
%}
%\label{fig:dens1}
%\end{figure}

\subsection{Representative elementary volume REV}

\begin{figure}[ht]
\begin{center}
\epsfig{file=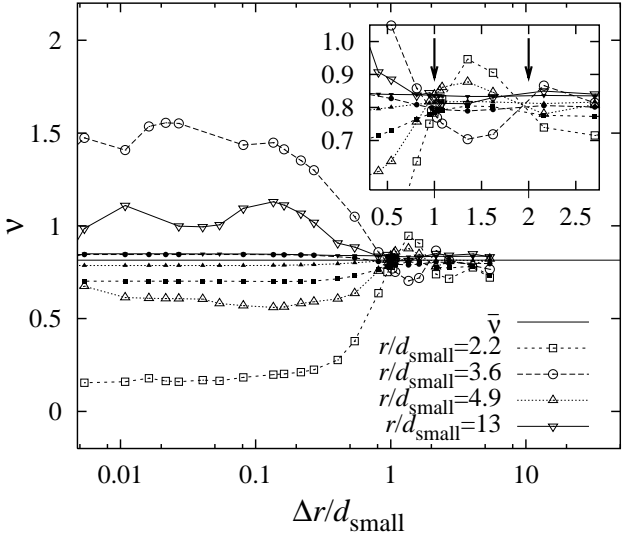,height=8.6cm,angle=-90}
\end{center}
\caption{Volume fraction $\nu$ at different distances $r$ from the inner 
ring, plotted against the width $\Delta r$ of the averaging ring
from simulation D, see table\ \protect\ref{tab:sim_details}. 
Note that the horizontal axis is logarithmic. The open symbols are results 
obtained with the particle-center method, the solid symbols are results from
the slicing method. The inset is a zoom into the large $\Delta r$ region,
and the arrows indicate the optimal width $\Delta r$ for the particle-center
method for which the results are almost independent of the averaging 
procedure.
}
\label{fig:deltar}
\end{figure}

An important question is, how does the result of
an averaging procedure depend on the size of the averaging 
volume $V$. We combine time- and space averaging, i.e.\ we average
over many snapshots and over rings of width $\Delta r$, so
that the remaining ``size'' of the averaging volume is the width of the 
rings $\Delta r$.  In Fig.~\ref{fig:deltar} data for $\nu$ at fixed
position $R=0.12$, 0.13, 0.14, and 0.20\,m, but obtained with different width 
$\Delta r$, are presented. The positions correspond to $r/d_{\rm small}\approx 2.2$,
3.6, 4.9, and 13, when made dimensionless with the diameter of the small
particles. Both the particle-center method (open symbols) 
and the slicing method (solid symbols) are almost identical for 
$\Delta r/d_{\rm small} \ge 2$, for the larger $\Delta r$ the averaging
volume can partially lie outside of the system. For very small 
$\Delta r/d_{\rm small} \le 0.1$ the different methods lead to strongly
differing results, however, the values in the limit $\Delta r
\rightarrow 0$ are consistent, i.e.~independent of $\Delta r$ besides
statistical fluctuations.  In the intermediate regime $0.1$ $<
\Delta r/d_{\rm small} < 2$, the particle-center method strongly varies,
while the slicing method shows a comparatively smooth variation.

Interestingly, all methods seem to collapse at $\Delta r_{\rm REV}$ 
$\approx d_{\rm small}$ (and twice this value), nearly
to the size of the majority of the particles.  For the examined
situations, we observe
that the particle-center and the slicing method lead to similar
results for $0.97 \le \Delta r_{\rm REV}/d_{\rm small} \le 1.03$.
This indicates that the systems (and measurements
of system quantities) are sensitive to a typical length scale, which is
here somewhat smaller than the mean particle size. When using this special
$\Delta r_{\rm REV}$ value, one has $B=20$ or $B=21$ binning intervals. 
The open question of this being a typical length scale that also occurs in 
systems with a broader size spectrum, cannot be answered with our setup,
due to the given particle-size ratio.

%\subsection{Effective volume fraction}
%
%Since   %as we will see later, 
%many of the properties in granulates 
%are related to interacting particles, it is useful to introduce 
%the effective volume fraction $\nu_c$ which only accounts for particles 
%with $c$ contacts or more
%\begin{equation}
%  \label{eq:nueffect}
%  \nu_c(r)=\langle \theta({\cal C}^p-c) \rangle
%          =\frac{1}{V} \sum_{p \in V} w_V^p V^p \theta({\cal C}^p-c) ~,
%\end{equation}
%with the step function $\theta(x)=1$ for $x \ge 0$ and $\theta(x)=0$
%otherwise and the number of contacts ${\cal C}^p$ of particle $p$.  
%In Fig.\ \ref{fig:dens2} different contact bearing volume fractions are 
%displayed as obtained from simulation D, see table\ 
%\protect\ref{tab:sim_details}. For $c=0$ the
%density fluctuations are weak, whereas the dilation in the shear
%zone is more evident for $c \ge 1$.
%\begin{figure}[ht]
%\begin{center}
%\epsfig{file=dens2.ps,height=7.5cm,angle=-90}
%\end{center}
%\caption{Effective volume fractions $\nu_c$ for different 
%$c=0$, 1, 2, and 3. Symbols correspond to $B_s=20$, the solid
%lines to $B_s=60$, and the dashed line to $B_c=20$.
%}
%\label{fig:dens2}
%\end{figure}
%As a last remark concerning density profiles, we note that the 
%volume fractions for the different simulations vary less than
%their fluctuations. 
%whereas the effective volume fractions are 
%clearly different from one simulation to the other.

\subsection{Time averaging}
\label{sec:time_av}

\begin{figure}[ht]
\begin{center}
\epsfig{file=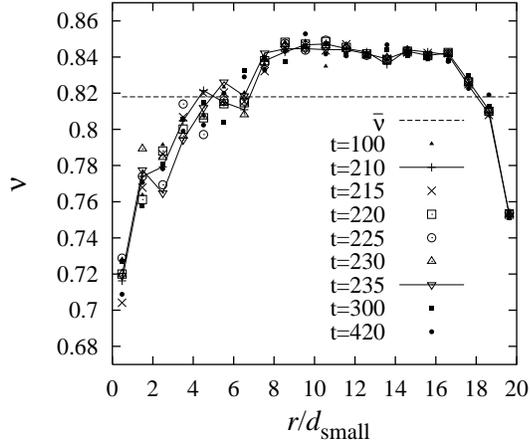,height=7.5cm,angle=-90}
\end{center}
\caption{Volume fraction $\nu$, plotted against the dimensionless
distance $r/d_{\rm small}$ from the inner ring, for different times
from simulation E.
Open symbols belong to six subsequent snapshots with $\Delta t=5$\,s, 
the small, solid symbols are snapshots after longer times.
}
\label{fig:times}
\end{figure}

In order to understand the fluctuations in the system over time,
and to test whether subsequent snapshots can be assumed to be
independent, the volume fraction $\nu$ is displayed for snapshots
at different averaging times in Fig.~\ref{fig:times}.

Changes in density are very weak and mostly occur in the dilated shear 
zone for small $r$. From one snapshot to the next, we frequently find, 
that the configuration in the outer part of the shear cell has not
changed, whereas a new configuration is found in the inner part.
Only after rather long times does the density change also in
the outer part.
Thus, simulation results in the outer part are subject to
stronger fluctuations because the average is 
performed over less independent configurations than in the
inner part.

\section{Macroscopic tensorial quantities}
\label{sec:macro}

In this section, the averaged, macroscopic tensorial quantities in
our model system are presented. The fabric tensor describes the contact
network, the stress tensor describes, in this study, the stress due
to normal forces, and the deformation gradient is a measure for the
corresponding elastic, reversible deformations. A more detailed 
analysis, where the tangential forces are also taken into
account, is in progress \cite{latzel99a} and some details in that
direction can be found in Refs.\ \cite{kruyt96,liao97b,kuhl99}.  
However,
we checked the influence of the tangential forces in our system
and found their effect to be always smaller than ten per-cent.

\subsection{Micro-mechanical fabric tensor}

\begin{figure}[ht]
\begin{center}
\epsfig{file=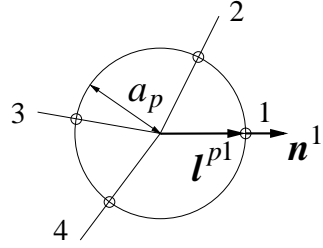,height=3.2cm,angle=0}
\end{center}
\caption{Schematic plot of a particle with radius $a$ and ${\cal C}^p=4$
contacts as indicated by the small circles. The branch vector
$\vec{l}^{pc}$ and the normal unit vector $\vec{n}^c$ are displayed
at contact $c=1$.}
\label{fig:nets}
\end{figure}

In assemblies of grains, the forces are transmitted from one particle
to the next only at the contacts of the particles.  In the general
case of non-spherical particles, a packing network is characterized by
the vectors connecting the centers of the particles and by the
particle-contact vectors. Furthermore, the local geometry of each
contact is important \cite{goddard98,cowin88}, see Fig.\ \ref{fig:nets}. 
In our case, with spherical
particles, the situation is simpler with respect to both the spherical
contact geometry and the fabric.

~\vspace{-1.3cm}\\
\subsubsection{Fabric tensor for one particle}

One quantity that describes the contact configuration of one particle to 
some extent, is the second order fabric tensor \cite{goddard98,cowin88}
\begin{equation}
\matrix{F}^p = \sum_{c=1}^{{\cal C}^p} \vec{n}^c \otimes \vec{n}^c ~,
\label{eq:fabric}
\end{equation}
where $\vec{n}^c$ is the unit normal vector at contact $c$ of particle
$p$ with ${\cal C}^p$ contacts. The symbol $\otimes$ denotes the 
dyadic product in this study. Other definitions of the fabric
use the so-called branch vector $\vec{l}^{pc}$ from the center of
particle $p$ to its contact $c$, however, the unit normal and the unit
branch vector are related by  $a_p \vec{n} = \vec{l}^{pc}$ in the case 
of spheres or disks, so that the definition
\begin{equation}
\matrix{F}^p = \frac{1}{a_p^2} \sum_{c=1}^{{\cal C}^p}
                                   \vec{l}^{pc} \otimes \vec{l}^{pc} ~.
\label{eq:fabric_ll}
\end{equation}
is identical to Eq.~(\ref{eq:fabric}).
\begin{figure*}[tb]
\epsfig{file=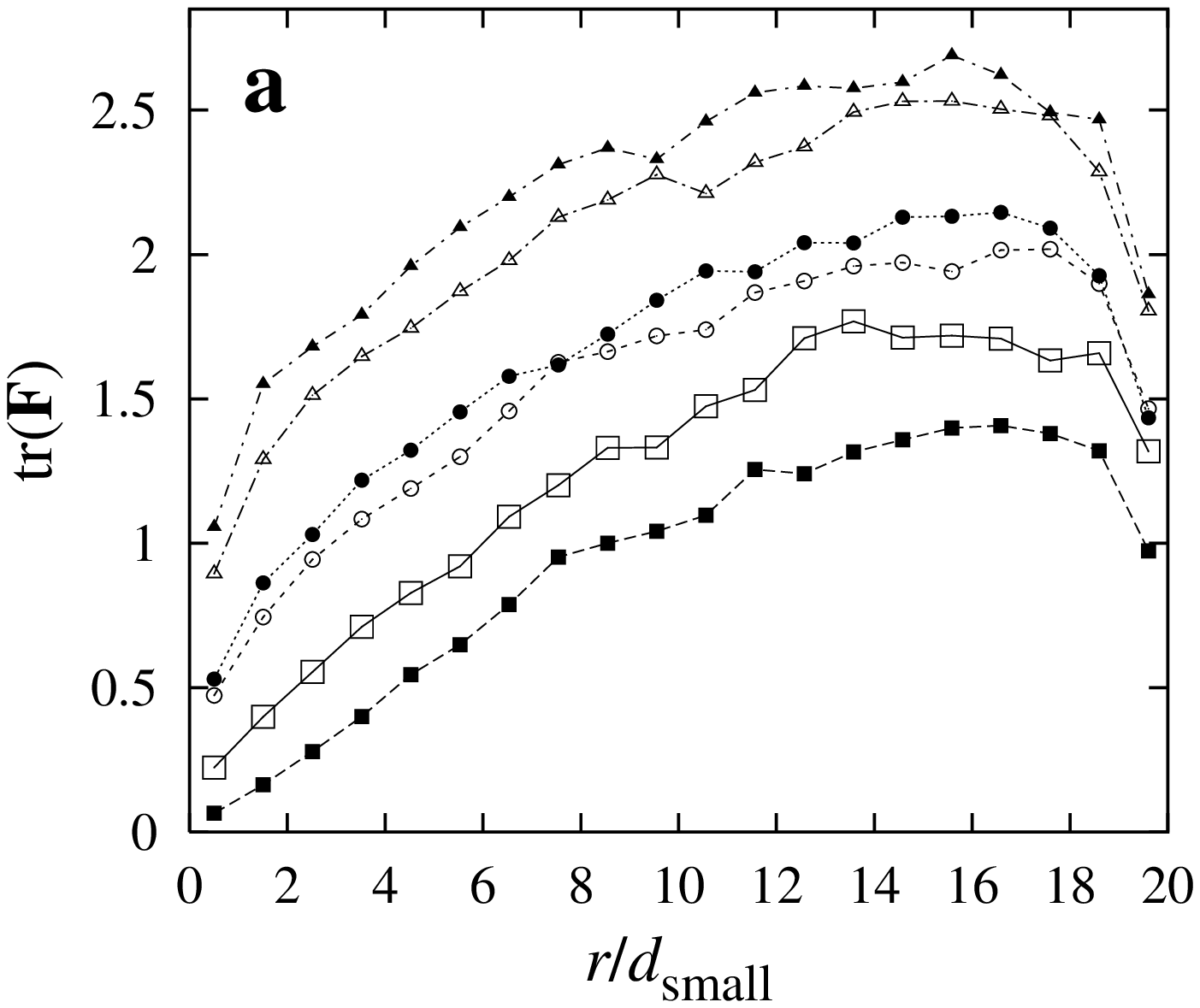,width=4.8cm,angle=-90} %\hfill
\epsfig{file=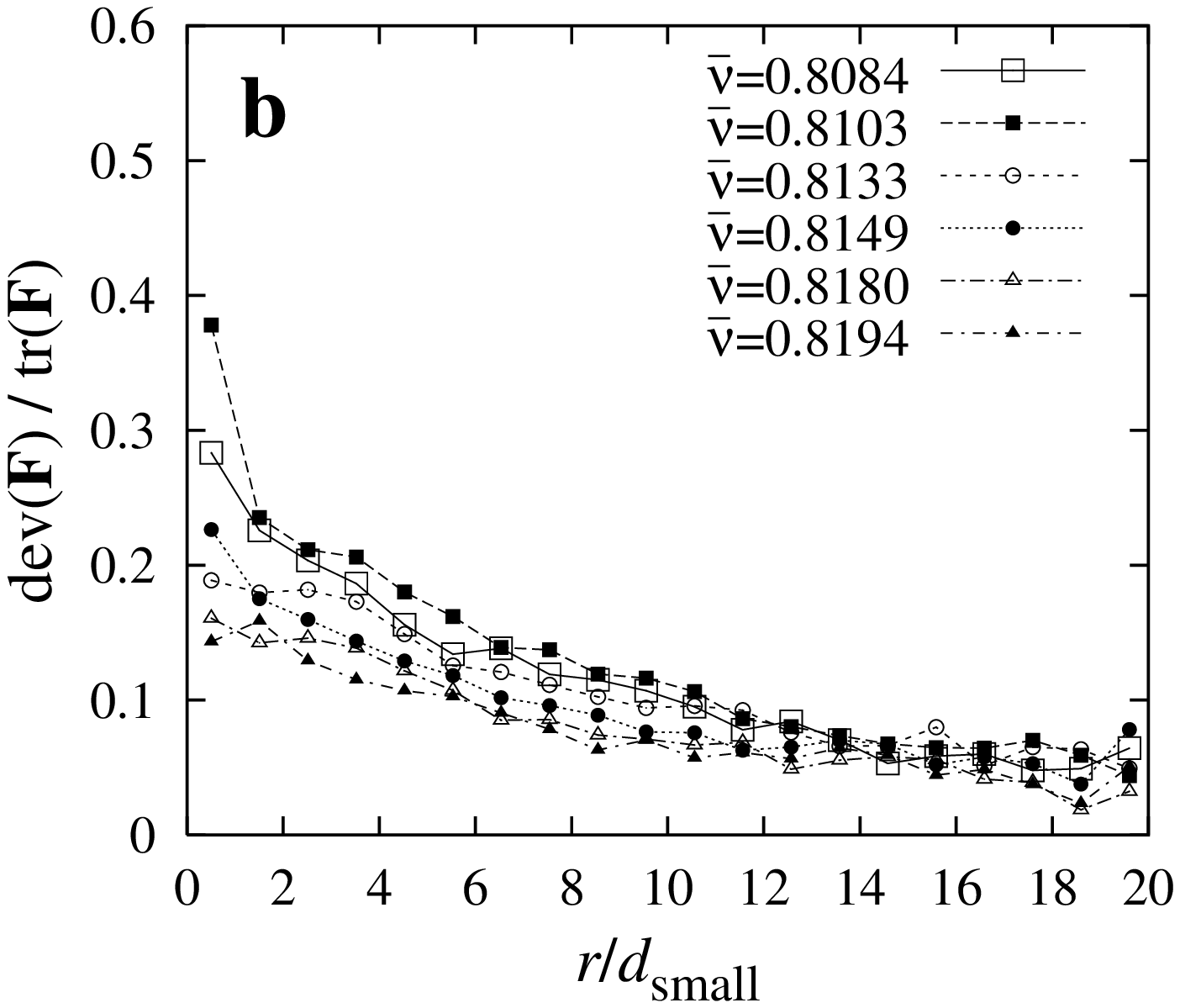,width=4.8cm,angle=-90} %\hfill
\hspace{-2.8cm}~
\epsfig{file=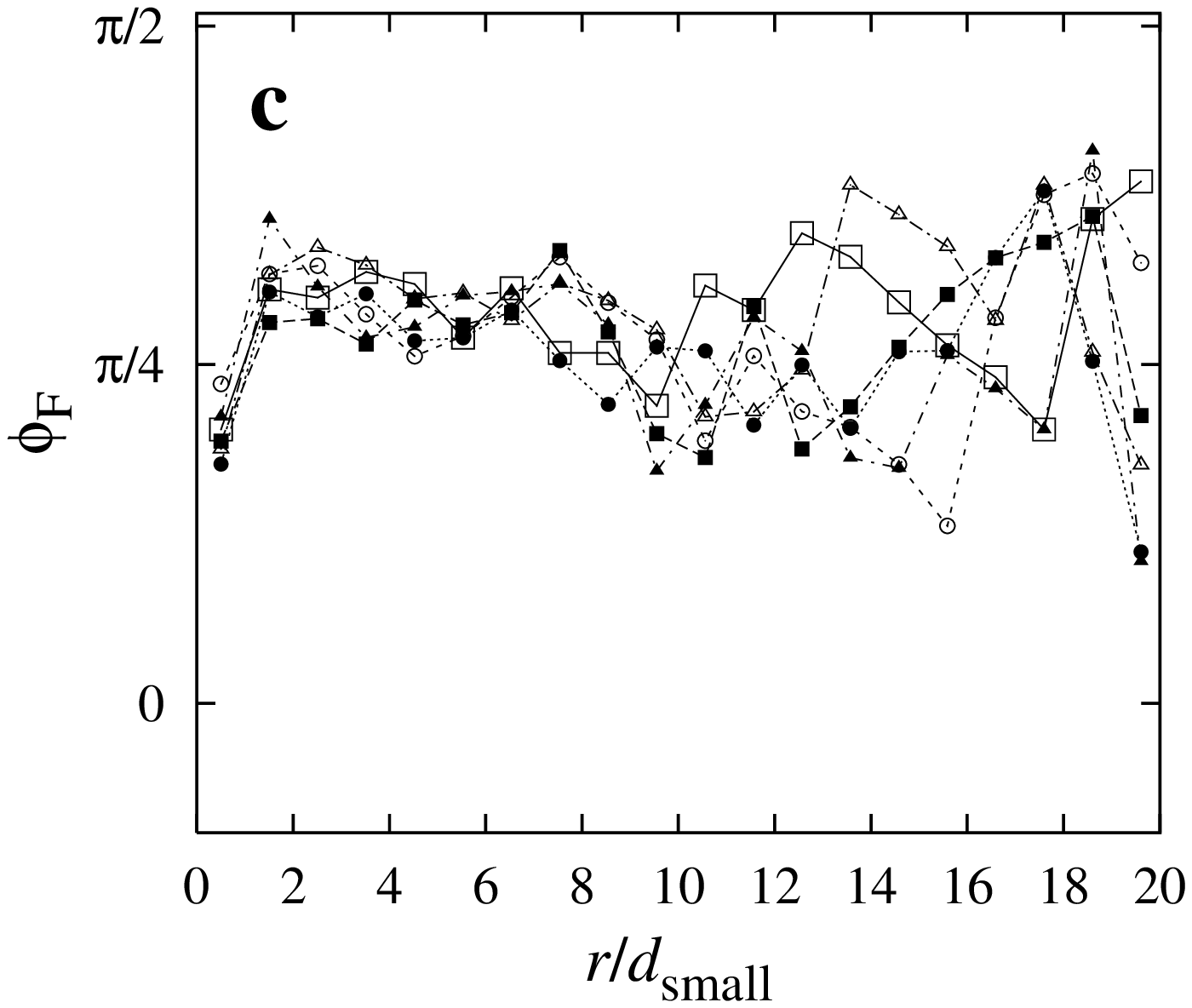,width=4.8cm,angle=-90} 
\caption{{\bf a-c} Plot of ({\bf a}) $F_V=\tr(\matrix{F})$,
({\bf b}) $F_D/F_V=\dev(\matrix{F})/\tr(\matrix{F})$, and
({\bf c}) $\phi_F$ against the dimensionless distance $r/d_{\rm small}$ 
from the inner ring.  The global volume fraction
is given in ({\bf b}) and valid also for ({\bf a}) and ({\bf c}).
}
\label{fig:fabric}
\end{figure*}
The fabric tensor in Eq.~(\ref{eq:fabric}) is symmetric by definition
and thus consists of up to three independent scalar quantities in two
dimensions. The first of them, the trace (or volumetric part) $F_V =
\tr(\matrix{F}^p) = (F_{\rm max}+F_{\rm min})$, is the number of
contacts of particle $p$, with the major and the minor eigenvalues
$F_{\rm max}$ and $F_{\rm min}$, respectively. One gets 
from Eqs.~(\ref{eq:fabric}) or (\ref{eq:fabric_ll}) the number 
of contacts of particle $p$
\begin{equation}
\tr(\matrix{F}^p) 
  = \sum_{c=1}^{{\cal C}^p} \tr(\vec{n}^c \otimes \vec{n}^c) 
  = {\cal C}^p ~, 
\label{eq:Cp}
\end{equation}
since the scalar product of $\vec{n}^c$ with itself is unity by 
definition.
The second scalar, the deviator $F_D = F_{\rm max}-F_{\rm min}$, 
accounts for the anisotropy of the contact network to first order,
and the third, the angle $\phi_F$, gives the orientation of 
the ``major eigenvector'', i.e.\ the eigenvector corresponding to 
$F_{\rm max}$, with respect to the radial direction.
In other words, the contact probability distribution is 
proportional to the function $[F_V + F_D \cos(2(\phi-\phi_F))]$
\cite{tsoungui98b,dubujet98,schollmann99}, when averaged
over many particles, an approximation which is not always
reasonable \cite{mehrabadi88}. 

\subsubsection{Averaged fabric tensor}

Assuming that all particles in $V$ contribute to 
the fabric with a weight of their volume $V^p$ one has
\begin{equation}
\matrix{F} = \langle \matrix{F}^p \rangle
           = \frac{1}{V} \sum_{p \in V} w_V^p V^p  
                         \sum_{c=1}^{{\cal C}^p}
                            \vec{n}^{c} \otimes \vec{n}^{c} ~.
\label{eq:vfabric}
\end{equation}
%We could imagine two possibilities for $V^p$: The first is to divide
%the volume in polygons with a Voronoi tessellation with only one particle 
%per polygon such that the polygons cover the whole volume. In that case,
%$V^p$ is the volume of the polygon that contains particle $p$.
%However, as discussed above, we will use the second possibility, 
%i.e.~we use the volume of particle $p$ so that $V^p=\pi a_p^2$. 
%Inserting our definition of $V^p$ into Eq.~(\ref{eq:vfabric}) 
%leads to the averaged fabric tensor 
%\begin{equation}
%\matrix{F} = \frac{\pi}{V} \sum_{p \in V} w_V^p
%               \sum_{c=1}^{{\cal C}^p} \vec{l}^{pc} \otimes \vec{l}^{pc} ~.
%\label{eq:vvfabric}
%\end{equation}
%
The values of $F_V$, $F_D/F_V$, and $\phi_F$, as obtained from our
simulations, are plotted in Figs.~\ref{fig:fabric}(a-c).
The trace of the fabric tensor and thus the mean number of 
contacts increases with increasing distance from the inner ring,
and is reduced in the vicinity of the walls. With increasing mean density,
the trace of $\matrix{F}$ is systematically increasing, while
the deviatoric fraction seems to decrease; this means that a denser
system is slightly more isotropic concerning the fabric.  The major 
eigendirection is tilted counterclockwise by somewhat more than $\pi/4$ 
from the radial outward direction, except for the innermost layer and for 
the strongly fluctuating outer region.

\begin{figure}[ht]
\begin{center}
 \epsfig{file=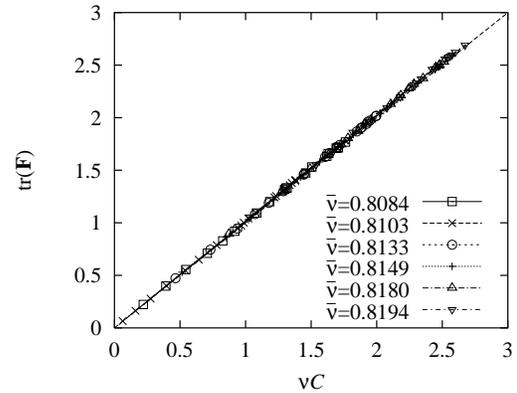,height=6.9cm,angle=-90} 
\end{center}
\caption{Trace of the fabric tensor plotted against $\nu {\cal C}$ 
  for different packing fractions $\bar \nu$. All the points collapse 
  nearby the identity curve (dashed line).}
\label{fig:trFtest}
\end{figure}

In analogy to the trace of the fabric for a single particle, the
trace of the averaged fabric is
\begin{equation}
\tr(\matrix{F}) = \frac{1}{V} \sum_{p \in V} w_V^p V^p {\cal C}^p 
                = \langle {\cal C}^p \rangle~,
\label{eq:trF}
\end{equation}
which, in the case of a regular, periodic contact network of almost 
identical particles with $a_p \simeq a$, reduces to the sum over all
of particles in $V$ with the prefactor $\nu$ defined in Eq.~(\ref{eq:nureal}).
Now, one has a relation between the coordination number, i.e.~the
mean number of contacts per particle 
${\cal C}={\cal C}(r)=\langle {\cal C}^p \rangle / \nu$, 
the volume fraction $\nu$, and the averaged fabric $\matrix{F}$, as a 
combination of Eqs.~(\ref{eq:nureal}) and (\ref{eq:trF}):
\begin{equation}
\tr({\matrix{F}}) \simeq \nu {\cal C} ~.
\label{eq:trFmono}
\end{equation}
As a test for the averaging procedure, we plot in Fig.~\ref{fig:trFtest} 
$\tr({\matrix{F}})$ against $\nu {\cal C}$ and obtain all 
data points from all simulations collapsing close to the identity curve.
For a theoretical derivation of the small (about 1 per-cent) deviation
due to the polydispersity, see Ref.\ \cite{tsoungui99}.

\subsection{Micro-mechanical stress tensor}
\label{sec:stress}

The micro-mechanical stress tensor is derived in a way similar
to Ref.\ \cite{kruyt96}.
For an arbitrary volume $V$ with surface $\partial V$, the mean
stress is defined as
\begin{equation}
\bsigma = \frac{1}{V} \int_V { dV'~ \bsigma'}~,
\end{equation}
where $\bsigma'=\bsigma'(\vec{x})$ is the position dependent local 
stress inside $V$ and $\vec{x}$ is the coordinate integrated over.
Note that $\bsigma'$ can be a strongly fluctuating function of
$\vec{x}$.

If the stress in the pore-space vanishes, the integral can be split 
into a sum over the stresses $\bsigma^p$, pre-averaged for particles 
$p$, with the respective center of mass vectors $\vec{x}^p$.
The mean stress is thus
\begin{equation}
\bsigma = \langle \bsigma \rangle
   = \frac{1}{V} \sum_{p \in V} w_V^p V^p \bsigma^p 
   = \frac{1}{V} \sum_{p \in V} w_V^p {\int_{V^p} { dV' \, \bsigma' }}~,
\label{eq:sum1p}
\end{equation}
which corresponds to a smearing out of $\bsigma'$ over the 
particles, with $\bsigma^p$, the average stress in particle
$p$, to be derived in the following.  

One can rewrite the transposed stress tensor so that
\begin{equation}
\bsigma^{\rm T} = \grad \vec{x} \cdot \bsigma^{\rm T}
                        = \div (\vec{x} \otimes \bsigma)
                         - \vec{x} \otimes \div \bsigma ~,
\label{eq:sigma_new}
\end{equation}
by introducing the unit tensor $\matrix{I} = \grad \vec{x}$,
and by applying the series rule in the first term on the right 
hand side. Note that the tensors $\bsigma$ on the right hand 
side are transposed with respect to the left hand side 
$\bsigma^{\rm T}$.  The transposed stress is used for the
following operations for the sake of simplicity, however, using 
the stress directly should lead to the same result.  For a more
detailed treatment, see Ref.\ \cite{latzel99a}. In static
equilibrium and in the absence of body forces, the term 
$\div \bsigma$ on the right hand side vanishes and $\bsigma$
is symmetric.

\subsubsection{Mean stress for one particle}

Using the definition of the transposed stress tensor in 
Eq.~(\ref{eq:sigma_new}) the integral over one particle
in Eq.~(\ref{eq:sum1p}) becomes
\begin{eqnarray}
({\bsigma^p})^{\rm T} & = & \frac{1}{V^p} \int_{V^p} dV'~ \div
  (\vec{x} \otimes \bsigma')\nonumber\\ 
& = &\frac{1}{V^p} \int_{\partial V^p}  { dS~ (\vec{x} \otimes
  \bsigma')\cdot \vec{n}}
\label{eq:intdv}
\end{eqnarray}
by application of Gauss' theorem. In Eq.~(\ref{eq:intdv}) $dS$ is the
surface element of $V^p$ on $\partial V^p$ and $\vec{n}$ is the
outwards normal unit vector. Using the definition of 
the stress vector $\vec{t} = \bsigma' \cdot \vec{n}$
one arrives at
\begin{equation}
({\bsigma^p})^{\rm T} = \frac{1}{V^p} \int_{\partial V^p} { dS~
  \vec{x} \otimes \vec{t} } ~.
\end{equation}
A force $\vec f^{c}$ acting at a 
contact $c$ with area $\delta s^c$ leads to a stress vector 
$\vec t^{c} = \vec f^{c} / \delta s^c$ -- even in the limit
of small contact area $\delta s^c \ll a_p$.
Here, we apply the simplifying assumption that
the force $\vec f^{c}$ is constant on the surface $\delta s^c$,
i.e.~we do not resolve any details at the contact.
Explicitly writing the surface element $dS$ as a sum, 
\begin{equation}
dS = \sum_{c=1}^{{\cal C}^p} \delta ( |{\vec x} - {\vec x}^c| ) ~ \delta s^c~,
\end{equation}
leads, after integration over the delta functions, to the transposed stress 
\begin{equation}
(\bsigma^p)^{\rm T} = \frac{1}{V^p} \sum_{c=1}^{{\cal C}^p}
\vec{x}^c \otimes \vec{f}^c ~.
\label{eq:sigmapT}
\end{equation}
Transposing Eq.~(\ref{eq:sigmapT}) leads to an exchange of 
$\vec{x}^c$ and $\vec{f}^c$,
and thus to the expression for the mean stress inside particle $p$:
\begin{equation}
\bsigma^p = \frac{1}{V^p} \sum_{c=1}^{{\cal C}^p} \vec{f}^c
\otimes \vec{x}^c ~,
\label{eq:sigmap}
\end{equation}
which is, at first glance, dependent on the frame of reference 
associated with the vector $\vec{x}^c$.
%
%\begin{figure}[ht]
%\begin{center}
%\epsfig{file=2p.eps,height=5.5cm,angle=0}
%\end{center}
%\caption{Schematic plot of two particles $p$ and $q$ with their
%common contact $c$.}
%\label{fig:2p}
%\end{figure}

Using vector addition, one has $\vec{x}^c = \vec{x}^p + \vec{l}^{pc}$,
with the position vector $\vec{x}^p$ of particle $p$ and the branch
vector $\vec{l}^{pc}$, pointing from the center of mass
of particle $p$ to contact $c$. Inserting this relation in Eq.~(\ref{eq:sigmap}) leads to
\begin{eqnarray}
\bsigma^p & = & \frac{1}{V^p} \left ( \sum_{c=1}^{{\cal C}^p}
  \vec{f}^c \right ) \otimes \vec{x}^p 
                    + \frac{1}{V^p} \sum_{c=1}^{{\cal C}^p} \vec{f}^c
                  \otimes \vec{l}^{pc}\nonumber \\ 
                  & = & \frac{1}{V^p} \sum_{c=1}^{{\cal C}^p} \vec{f}^c
                  \otimes \vec{l}^{pc} ~,
\label{eq:sigmalpc}
\end{eqnarray}
since the first sum vanishes in static equilibrium, where
\begin{equation}
\sum_{c=1}^{{\cal C}^p} \vec{f}^c = \vec{0} ~.
\label{eq:equif}
\end{equation}

\subsubsection{Averaged stress tensor}

Inserting Eq.~(\ref{eq:sigmalpc})  in Eq.~(\ref{eq:sum1p}) gives a 
double sum over all particles with center inside the averaging volume $V$
and all their contacts
\begin{equation}
\bsigma = \langle \bsigma^p \rangle 
        = \frac{1}{V} \sum_{p \in V} w_V^p
          \sum_{c=1}^{{\cal C}^p} \vec{f}^{c} \otimes \vec{l}^{pc} ~.
\label{eq:sigma_av}
\end{equation}
In other words, stress is pre-averaged over all particles 
and then averaged over $V$.

Inserting  Eq.~(\ref{eq:sigmap}) in Eq.~(\ref{eq:sum1p}) 
leads to the mathematically identical sum
\begin{equation}
\bsigma = \frac{1}{V} \sum_{p \in V} w_V^p
         \sum_{c=1}^{{\cal C}^p} \vec{f}^{c} \otimes \vec{x}^c ~.
\label{eq:sigmapc}
\end{equation}
This expression can be transformed into a sum over all contacts $c(p,q)\in V$,
where at least one of the participating particles $p$ and $q$ lies inside the 
volume $V$. The force $\vec{f}^c=\vec{f}^{pq}=-\vec{f}^{qp}$,  
acting at contact $c$, from particle $q$ on $p$ is equal, but opposite
in direction to the 
force acting at the same contact from particle $p$ on $q$, so that  
\begin{eqnarray}
\bsigma & = & \frac{1}{V} \sum_{c \in V} 
                 ( w_V^p \vec{f}^{pq} + w_V^q \vec{f}^{qp} )\otimes \vec{x}^c 
                                                                       \nonumber \\
%        & = & \frac{1}{V} \sum_{c \in V} 
%                 ( w_V^p - w_V^q ) \, \vec{f}^{c} \otimes \vec{x}^c 
                                                                       \nonumber \\
        & = & \frac{1}{V} \sum_{c \in \partial V} 
                 ( w_V^p - w_V^q ) \, \vec{f}^{c} \otimes \vec{x}^c ~,
\label{eq:sigmax}
\end{eqnarray}
since $w_V^p - w_V^q=0$ when both particles are completely inside $V$.
Note that every contact is visited once in Eq.~(\ref{eq:sigmax}) but 
twice in Eqs.~(\ref{eq:sigma_av}) and (\ref{eq:sigmapc}).
The expression $c \in \partial V$ means for the particle-center 
averaging method that contacts contribute only if one particle is inside
$V$ while the other is outside.  In the framework of the slicing method,
contacts contribute if at least one particle is cut by $\partial V$
so that $w_V^p - w_V^q \ne 0$.

Given an arbitrary averaging volume $V$, and all the information
about the forces at all contacts from the discrete element simulations,
it is obvious that -- from the technical point of view --
the sum in Eq.~(\ref{eq:sigma_av})
% is straightforwardly performed,
can be done relatively easily, whereas the sum in
Eq.~(\ref{eq:sigmax}) requires the identification of the
contacts at the ``surface'' of $V$, before the possibly shorter
summation can be performed. Note that both expressions
Eq.~(\ref{eq:sigma_av}) and Eq.~(\ref{eq:sigmax}) have been used by
different authors, see e.g.~\cite{kruyt96,chang88,cundall82}.

\begin{figure*}[bt]
\epsfig{file=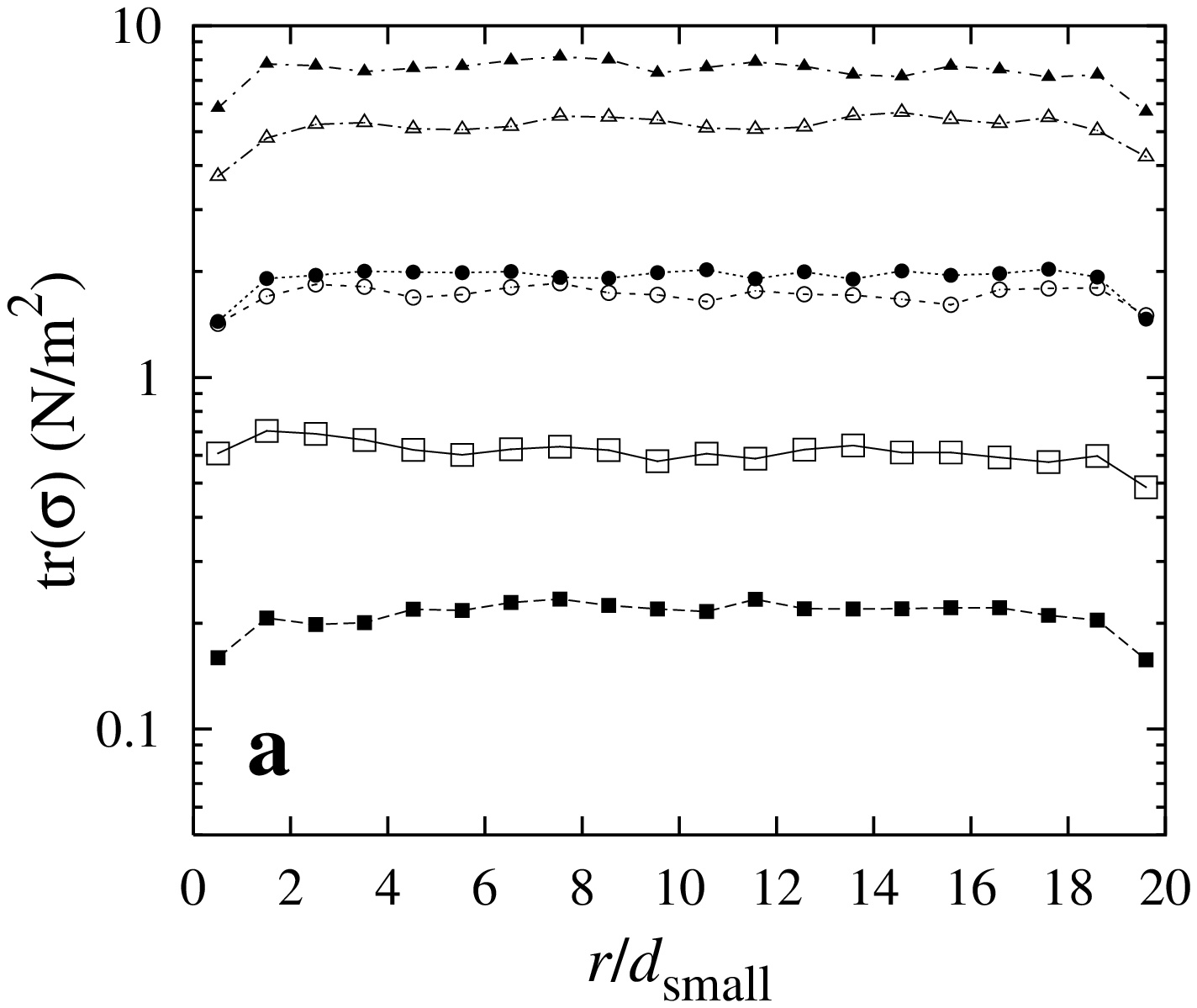,width=4.8cm,angle=-90} %\hfill
\epsfig{file=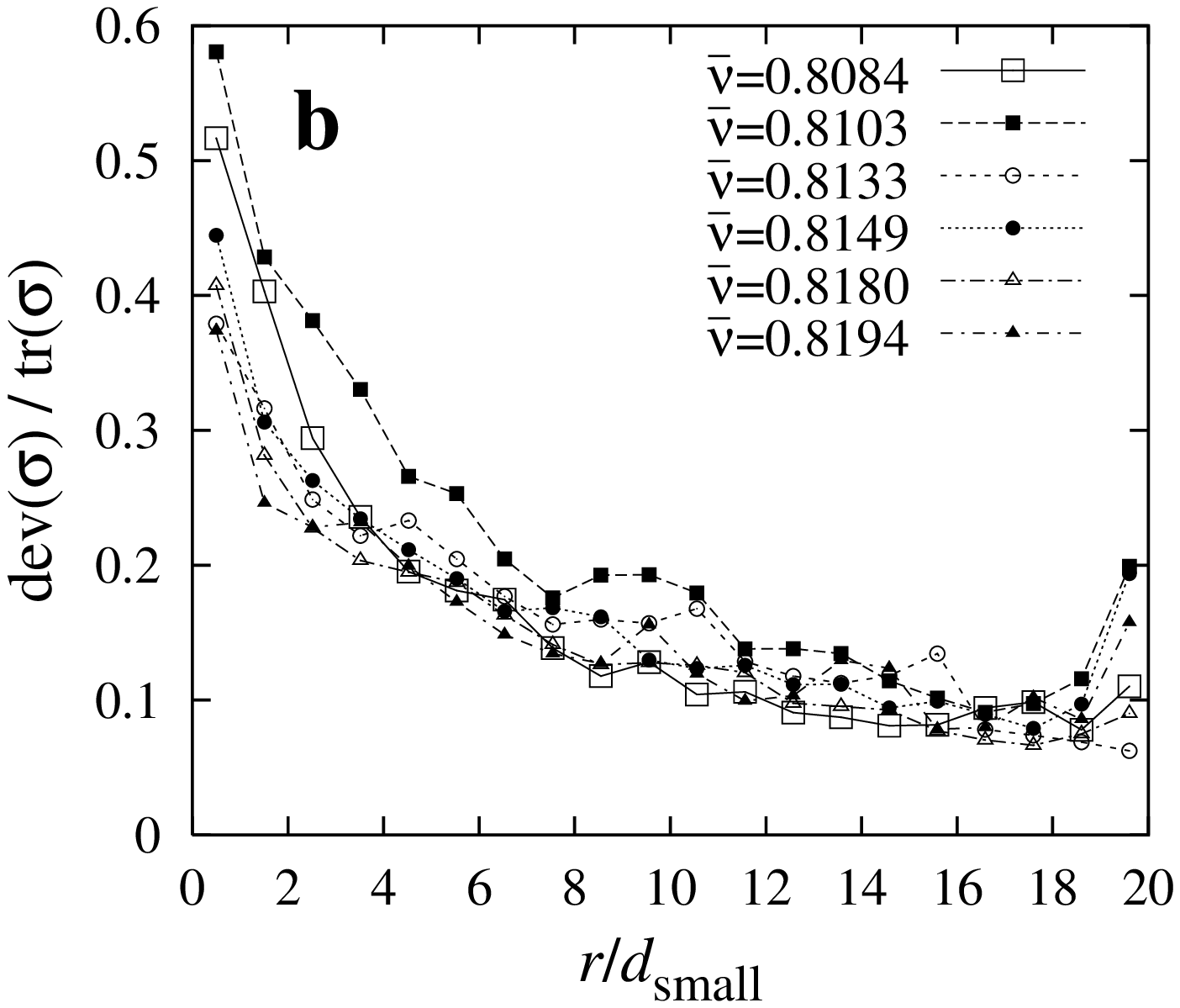,width=4.8cm,angle=-90} %\hfill
\hspace{-2.8cm}~
\epsfig{file=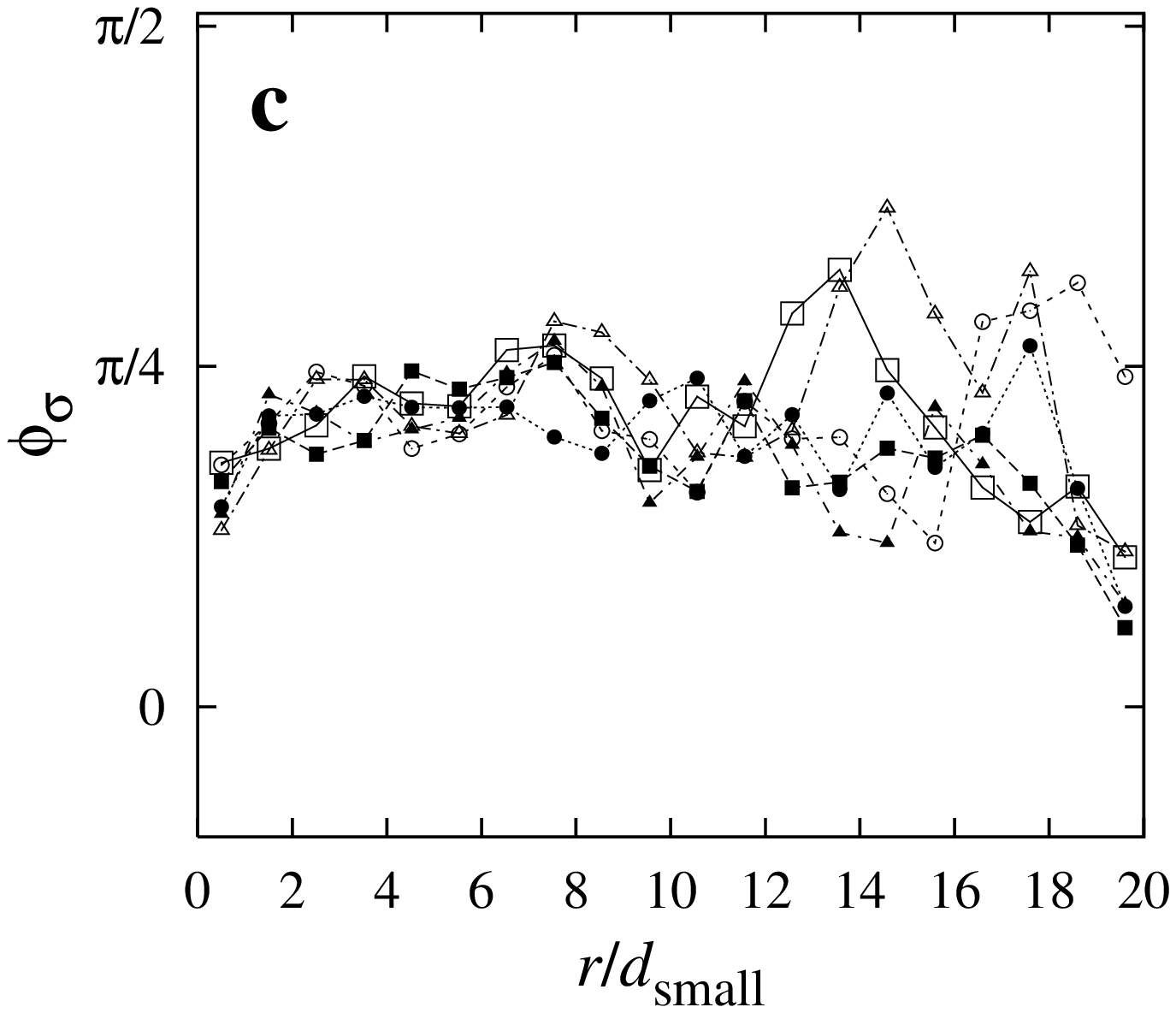,width=4.8cm,angle=-90} 
\caption{{\bf a-c} Plot of ({\bf a}) $\sigma_V=\tr(\bsigma)$,
({\bf b}) $\sigma_D/\sigma_V=\dev(\bsigma)/\tr(\bsigma)$, and
({\bf c}) $\phi_\sigma$ against the dimensionless distance $r/d_{\rm small}$ 
from the inner ring, as obtained with the slicing method. The global volume 
fraction is given in ({\bf b}) and valid also for ({\bf a}) and ({\bf c}).
}
\label{fig:stress}
\end{figure*}

\begin{figure*}[htb]
\epsfig{file=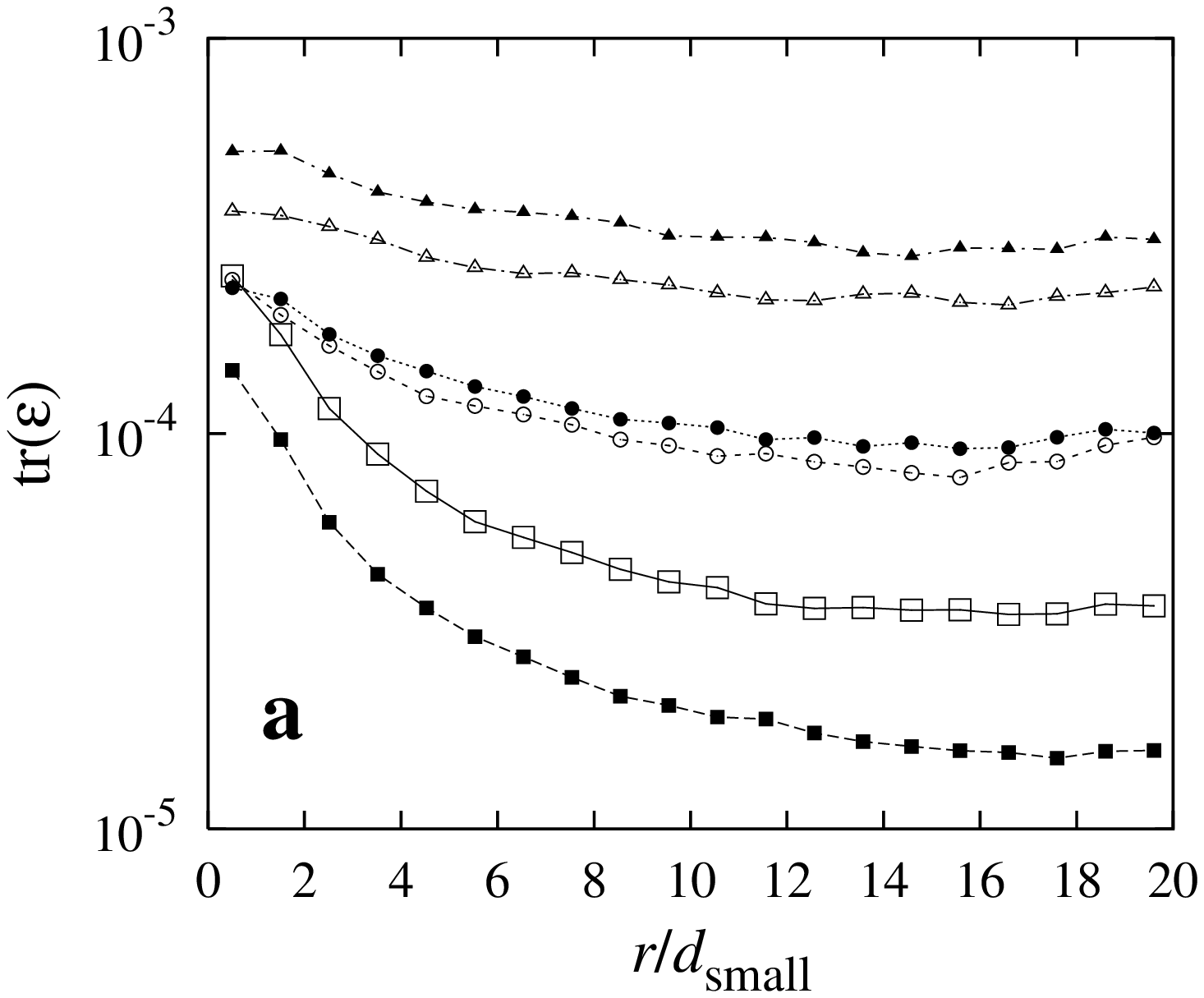,width=4.8cm,angle=-90} %\hfill
\epsfig{file=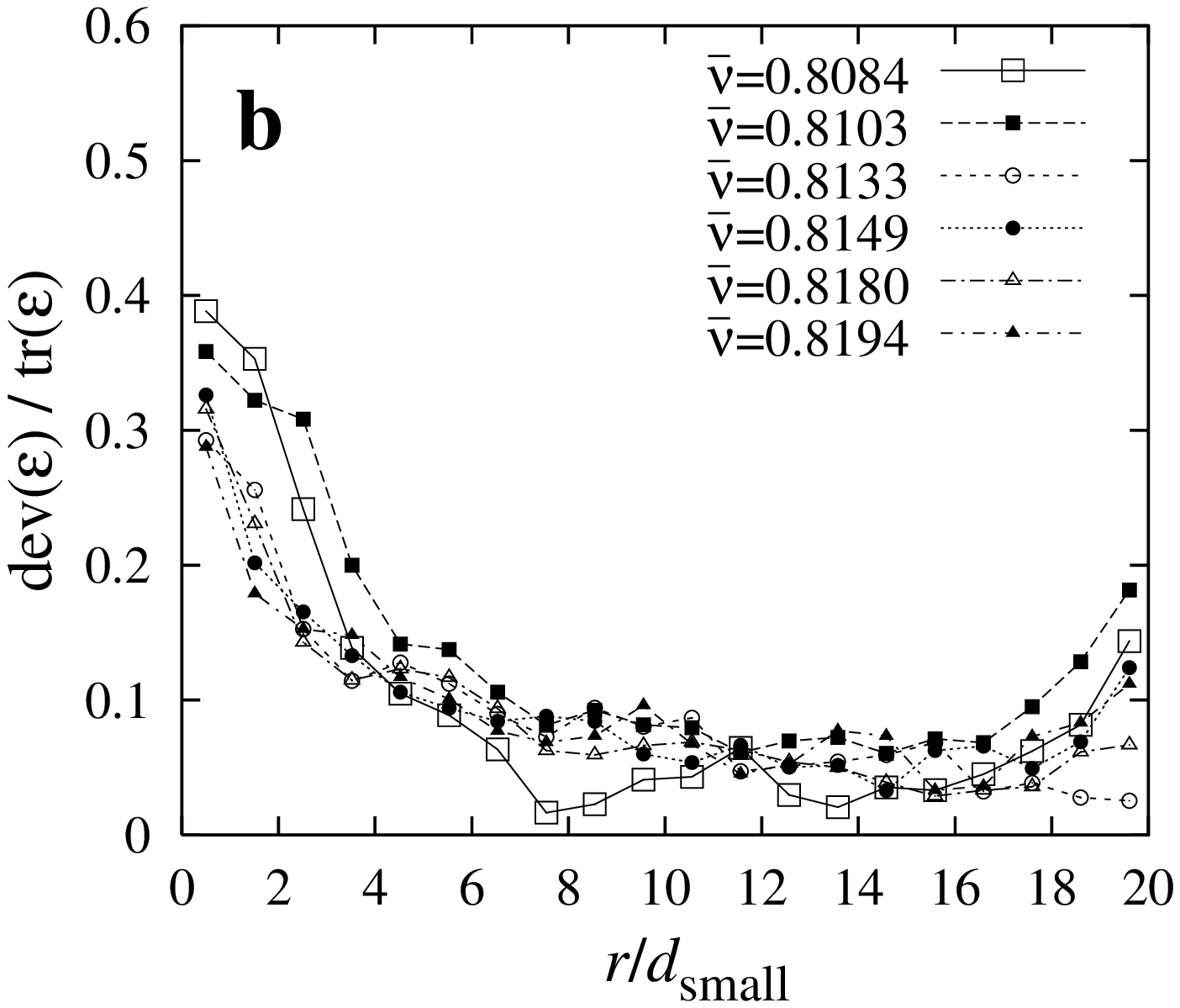,width=4.8cm,angle=-90} %\hfill
\hspace{-2.8cm}~
\epsfig{file=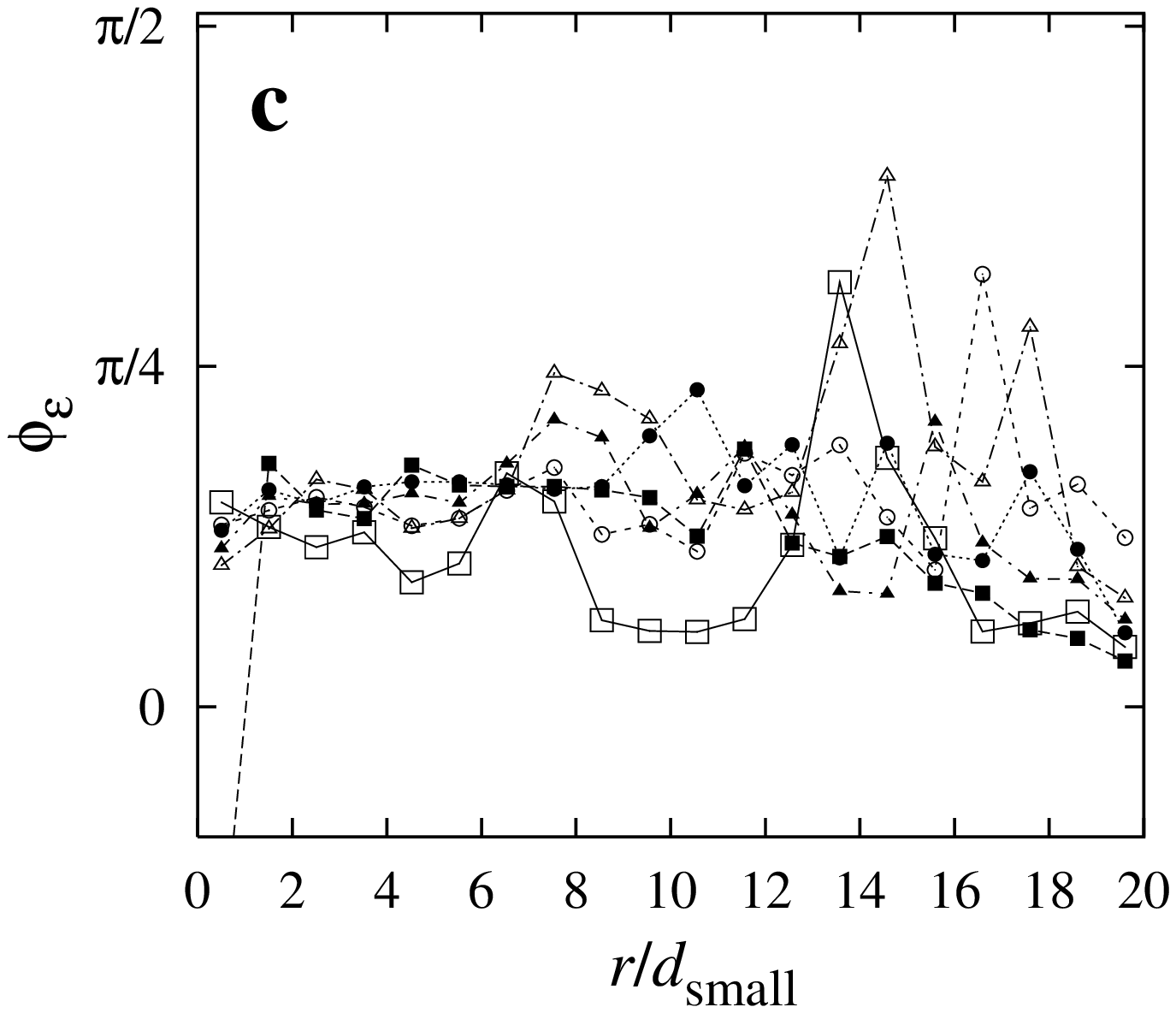,width=4.8cm,angle=-90} 
\caption{{\bf a-c} Plot of ({\bf a}) $\epsilon_V=\tr(\bvarepsilon)$,
({\bf b}) $\epsilon_D/\epsilon_V=\dev(\bvarepsilon)/\tr(\bvarepsilon)$,
and ({\bf c}) $\phi_\epsilon$ against the dimensionless distance 
$r/d_{\rm small}$ from the inner ring.  The global volume fraction
is given in ({\bf b}) and valid also for ({\bf a}) and ({\bf c}).
}
\label{fig:defgrd}
\end{figure*}

The values of the volumetric stress $\sigma_V=\tr(\bsigma)$, the
deviatoric fraction $\sigma_D/\sigma_V=\dev(\bsigma)/\tr(\bsigma)$,
as obtained from our simulations, are plotted in Figs.~\ref{fig:stress}(a-b).
The volumetric stress is constant, besides fluctuations, whereas the 
deviatoric fraction is rather large in the shear-zone and decays with 
increasing distance from the inner ring, like the fabric. Also, the
deviatoric fraction of the stress appears slightly reduced in magnitude 
for increasing global density. Note that the stress varies
over almost two orders of magnitude, while the mean density is
changed only weakly from $\bar \nu=0.8084$ to $0.8194$, see
table~\ref{tab:sim_details}. In Fig.~\ref{fig:stress}c $\phi_\sigma$
describes the angle which the principal axis of the stress tensor is tilted
from the outward direction. The stress tensor is, on average, rotated
counter clockwise by somewhat less than $\pi/4$ from the outward
direction. It is only in the outermost part that strong fluctuations
exist around the mean. We now provide evidence that the fabric and stress are
{\em not} co-linear.

\subsection{Mean strain tensor}

To achieve the material properties of a granular ensemble one is
interested in the stress-strain relationship of the material. 
One of the simplest techniques used, is the application of 
``Voigt's hypothesis'' which assumes that the strain is uniform 
and that every particle displacement conforms to the mean 
displacement field \cite{liao97}. Thus, the expected displacement 
at contact $c$ of particle $p$, relative to the force free situation, 
and due to the mean displacement gradient $\bvarepsilon$, is
$\bvarepsilon \cdot \vec{l}^{pc}$,
with $\vec{l}^{pc}=a_p \vec{n}^c$, and the mean contact 
deformation $\delta$.  Note that the linear, symmetric strain 
$\matrix{\bepsilon}=\frac{1}{2} (\bvarepsilon+\bvarepsilon^{\rm T})$
is not identical to the displacement gradient, in general.
In our study, we follow the approach of Liao et al. \cite{liao97b},
who used $\vec{l}^{pq}$ instead of $\vec{l}^{pc}$,
and assume that the actual displacement field does not coincide with,
but fluctuates about the mean displacement field. The difference
between the actual displacement $\vec{\Delta}^{pc}$ and the expected 
displacement is
\begin{equation}
\vec{\chi}^{pc}=\bvarepsilon \cdot \vec{l}^{pc}- \vec{\Delta}^{pc} ~.
\label{eq:diff_main}
\end{equation}
The actual displacement is directly related to the simulations
via $\vec{\Delta}^{pc}=\delta^c \vec{n}^{c}$.

If one assumes that the mean displacement field best approximates
the actual displacement, one can apply a ``least square fit'' to the 
total fluctuations 
\begin{equation}
S=\frac{1}{V} \sum_{p \in V} w_V^p \sum_{c=1}^{{\cal C}^p} 
  (\vec{\chi}^{pc})^2 ~,
\label{eq:Ssum}
\end{equation}
by minimizing $S$ with respect to the mean displacement gradient
so that
\begin{eqnarray}
\frac{\partial S}{\partial \bvarepsilon} & \stackrel{!}{=} & {\matrix 0} \\
& = & \frac{2}{V} \sum_{p \in V} w_V^p
      \sum_{c=1}^{{\cal C}^p} (\bvarepsilon\cdot\vec{l}^{pc}-\vec{\Delta}^{pc}) 
      \cdot \frac{\partial}{\partial \bvarepsilon}
     (\bvarepsilon\cdot\vec{l}^{pc}-\vec{\Delta}^{pc}) \nonumber ~.
\end{eqnarray}
These four equations for the four components of $\bvarepsilon$ in 2D 
can be transformed into a relation for the mean displacement tensor as 
a function of the contact displacements and the branch vectors.
By assuming that $\partial \vec{\Delta}^{pc}/\partial \bvarepsilon = {\vec 0}$,
the expression ``$\cdot \frac{\partial}{\partial \bvarepsilon}
(\bvarepsilon\cdot\vec{l}^{pc})$'' becomes a dyadic product 
``$\otimes \vec{l}^{pc}$'', as can be seen by writing down the equation 
in index notation. 
Extracting $\bvarepsilon$ from the sum (what leaves $\vec{l}^{pc}
\otimes \vec{l}^{pc}$, the core of the fabric, in the first term) and multiplying 
the equation with $\matrix{A}=\matrix{F}^{-1}$, the inverse of the fabric, see
Eq.~(\ref{eq:vfabric}), we find that
\begin{equation}
\bvarepsilon = \frac{\pi}{V} \left ( 
           \sum_{p \in V} w_V^p \sum_{c=1}^{{\cal C}^p}
           \vec{\Delta}^{pc} \otimes \vec{l}^{pc} 
     \right ) \cdot \matrix{A}  ~.
\label{eq:msdef}
\end{equation}
%By extracting $a_p$ and $\delta^c$ from the vectors in the inner sum,
%Eq.\ \ref{eq:msdef} can be transformed to 
%\begin{equation}
%\bvarepsilon = \frac{1}{V}
%           \sum_{p \in V} w_V^p V^p \sum_{c=1}^{{\cal C}^p}
%           \left (
%           \frac{\delta^c}{a_p} \, \vec{n}^{c} \otimes \vec{n}^{c} 
%     \cdot \matrix{A}  \right ) ~,
%\label{eq:msdef2}
%\end{equation}
%with the object in brackets $\matrix{E}^c=(\ldots)$, a fabric-weighted,
%directed deformation at contact.
 
The values of $\epsilon_V$, $\epsilon_D/\epsilon_V$, and $\phi_\epsilon$,
as obtained from our simulations, are plotted in Figs.~\ref{fig:defgrd}(a-c).
The elastic, volumetric deformation gradient of the granulate is 
localized in the shear zone and the effect is stronger in the less
dense systems. Due to dilation it is easier to compress the dilute
material closer to the inner ring compared to the outer
part. Like the fabric and stress, the strain also becomes more isotropic
with increasing mean density.

\section{Results}
\label{sec:results}

\begin{figure}[b]
\begin{center}
\epsfig{file=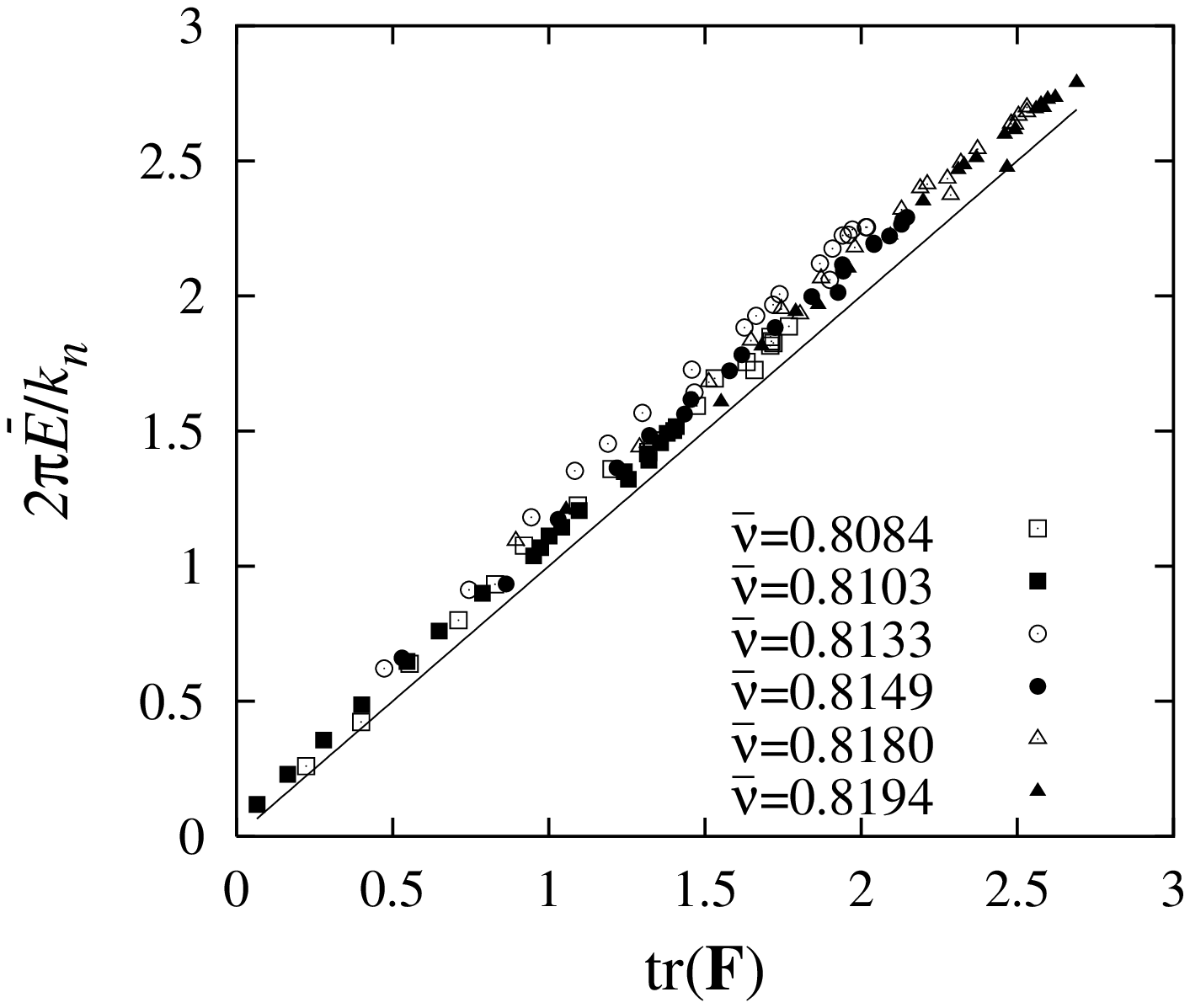,width=5.5cm,angle=-90}
\end{center}
\caption{Granulate stiffness 
$2 \pi \bar E/k_n=\tr(\bsigma)/\tr(\bvarepsilon)$,
plotted against $\tr(\matrix{F})$ from all simulations.
}
\label{fig:macro1}
\end{figure}

\begin{figure}[b]
\begin{center}
\epsfig{file=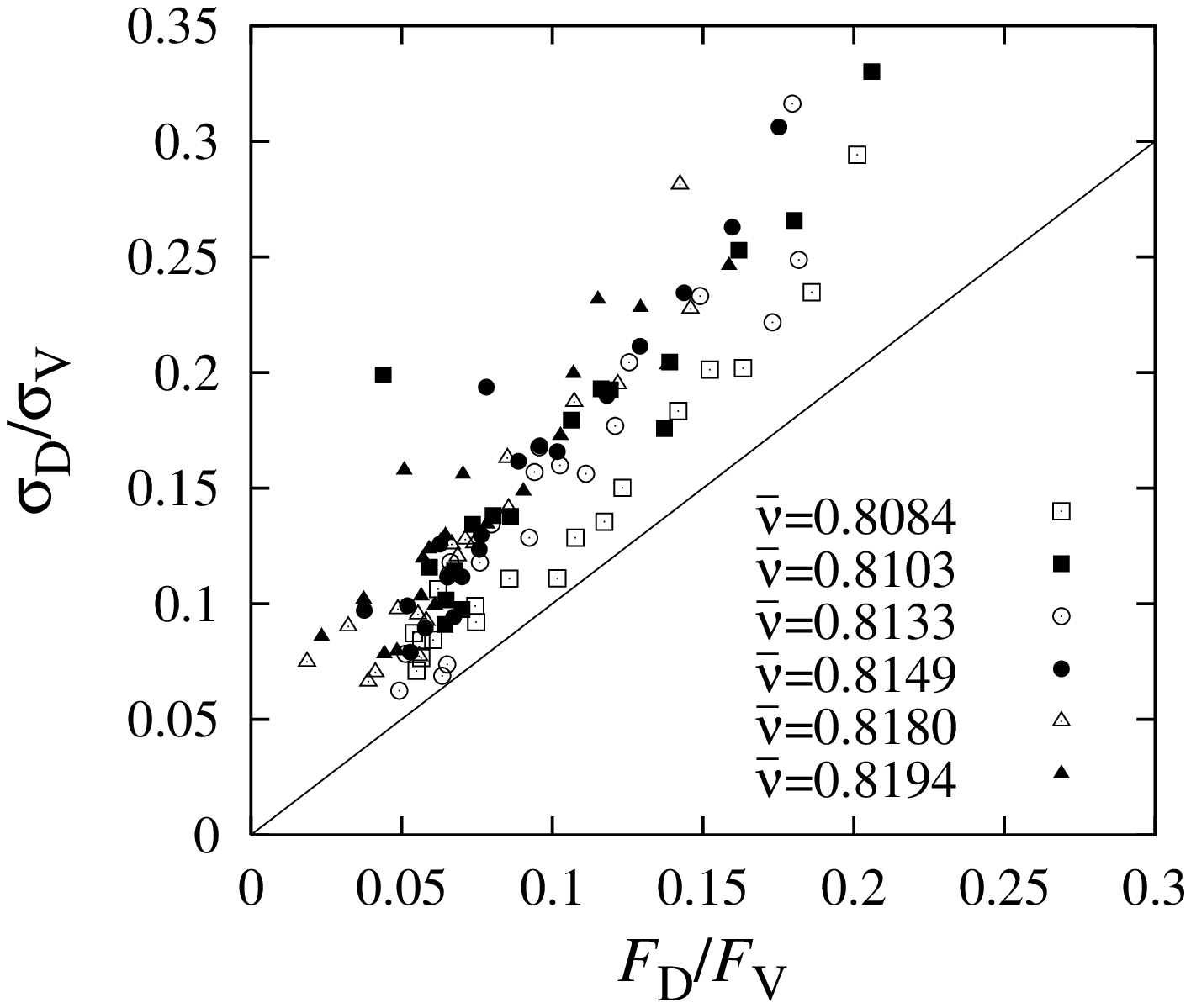,width=5.5cm,angle=-90}
\end{center}
\caption{Deviatoric fraction $\dev(\bsigma)/\tr(\bsigma)$
plotted against $\dev(\matrix{F})/\tr(\matrix{F})$ from all 
simulations.
}
\label{fig:macro11}
\end{figure}

At first, we compute the mean-field expectation values for $\bsigma$ and
$\bvarepsilon$, in order to get a rough estimate for the orders of magnitude
of the following results.
Replacing, in Eq.~(\ref{eq:sigma_av}), $f^c$ by its mean
$\bar f = k_n \bar \delta$, $a_p$ by $a = \bar a$, and $\vec{l}^{pc}$ by 
$a \vec{n}^c$, one gets 
\begin{equation}
\bar{\bsigma}= (k_n \bar \delta / \pi a) \,\matrix{F} ~.
\label{eq:barsigma}
\end{equation}
Performing some similar replacements in Eq.~(\ref{eq:msdef}), 
leads to 
\begin{equation}
\bar{\bvarepsilon}=(\bar \delta/a) \, \matrix{I}  ~,
\label{eq:bareps1}
\end{equation}
or
\begin{equation}
\bar{\bvarepsilon}=(\pi/k_n) \, \bsigma \cdot \matrix{A} ~.
\label{eq:bareps2}
\end{equation}
The material stiffness, $\bar{E}$, can be defined as the ratio of the
volumetric parts of stress and strain, so that one obtains from 
Eq.~(\ref{eq:barsigma}) and (\ref{eq:bareps1})
\begin{equation}
\bar E = (k_n/2\pi) \, \tr(\matrix{F}) ~.
\end{equation}
In Fig.~\ref{fig:macro1} the rescaled stiffness of the granulate
is plotted against the trace of the fabric for all simulations.
Note that all data collapse almost on a line, but the mean-field
value underestimates the simulation data by a few per-cent.
Simulation data for different $k_n$ and even data from simulations
with neither bottom- nor tangential friction collapse with the 
data for fixed $k_n$ and different volume fractions, shown here.
The deviatoric fraction of $\bsigma$ is plotted in 
Fig.~\ref{fig:macro11} against the deviatoric fraction
of $\matrix{F}$. Here, the data are strongly underestimated 
by the mean-field result in Eq.~(\ref{eq:barsigma}).
A similar plot for the deviatoric fraction of the deformation
gradient against $F_D/F_V$ shows a gathering of the data close
to the identity curve, in disagreement with Eq.~(\ref{eq:bareps1}).
Thus, we conclude that the deviatoric parts of stress, deformation
gradient and fabric are interconnected in a more complicated way 
than suggested by the simple mean field estimates 
\cite{kruyt96,liao97,luding99b,luding99c}.

\begin{figure}[t]
\begin{center}
\epsfig{file=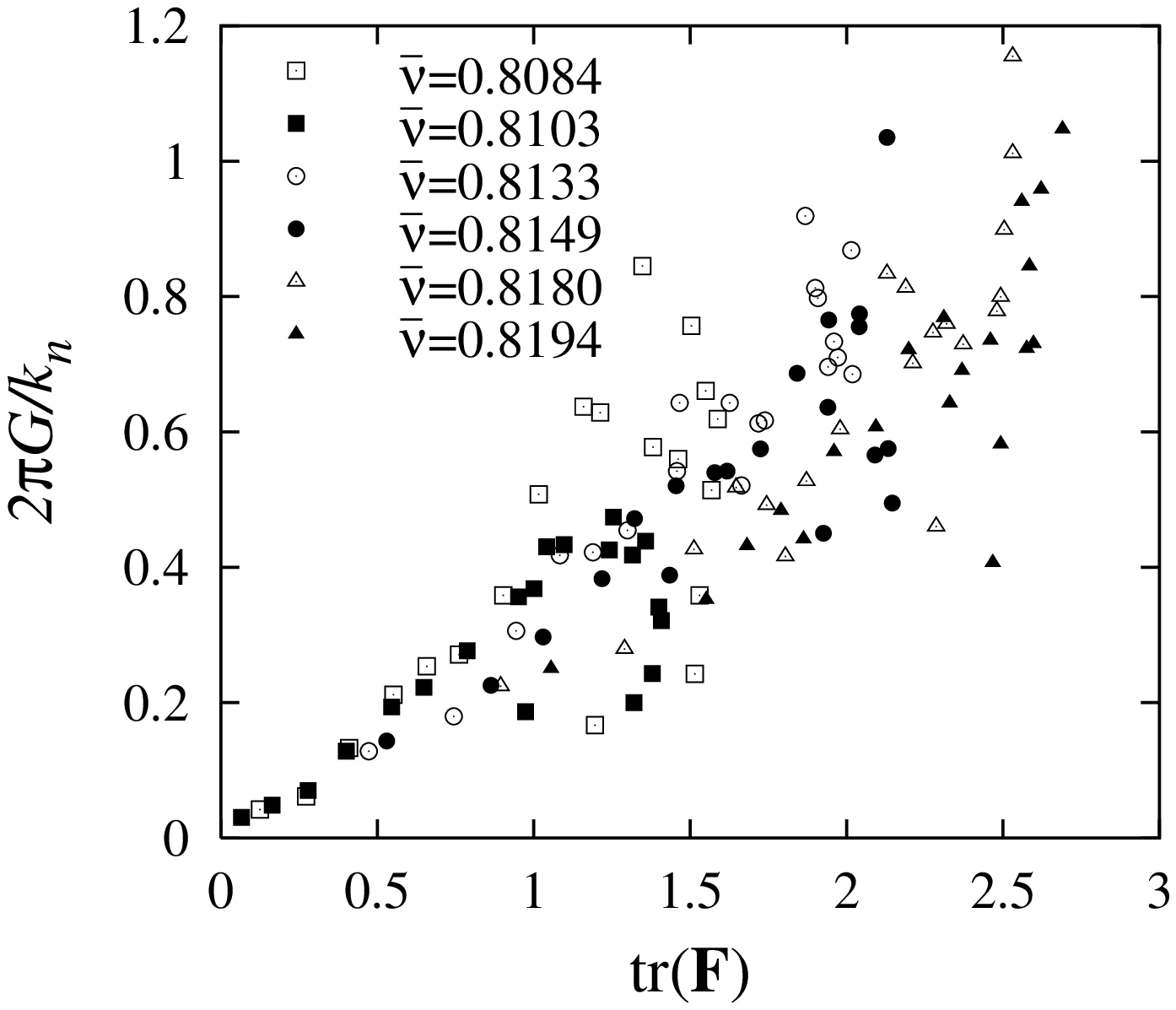,width=5.5cm,angle=-90}
\end{center}
\caption{Scaled granulate shear resistance 
$G=\dev(\bsigma)/\dev(\bvarepsilon)$
plotted against $\tr(\matrix{F})$ from all simulations.
}
\label{fig:macro2}
\end{figure}

In Fig.~\ref{fig:macro2} the ratio of the deviatoric parts
of stress and strain is plotted against the trace of the fabric.
Like the material stiffness, both quantities are proportional,
only $G$ shows much stronger fluctuations and has a proportionality
factor of about $1/3$. We did not use the traditional definition
of the shear modulus \cite{kruyt96}, since our tensors are not
co-linear as shown below.

\begin{figure}[b]
\begin{center} ~\hspace{-3.6cm}
\epsfig{file=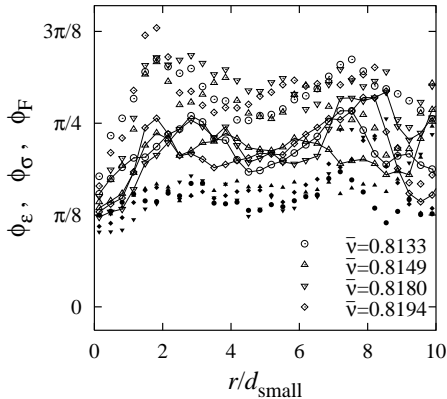,width=5.5cm,angle=-90} \\
\end{center}
\caption{Orientation of the tensors $\matrix{F}$, $\bsigma$, and $\bvarepsilon$
plotted against $r/d_{\rm small}$ from simulations C, D, E, and F with $B_s=60$. The 
data are shown in the range $0 \le r/d_{\rm small} \le 10$ only. Open symbols are
fabric, open symbols connected by lines are stress, and solid symbols are
strain, the mean densities corresponding to the symbols is given in the plot.
}
\label{fig:macro3}
\end{figure}

In Fig.~\ref{fig:macro3} the orientations of the tensors are plotted
for some simulations and in the inner part of the shear cell. In the 
outer part, the deviatoric fraction is usually around 10 per-cent, 
i.e.~so small that the orientations become noisier. Simulations A
and B are skipped here, because the data of the orientations are rather 
noisy due to the low density which leads to intermittent behavior with 
strong fluctuations. We observe that all orientation angles $\phi$
show the same qualitative behavior, however, the fabric is tilted
more than the stress which in turn is tilted more than the 
deformation gradient, where the orientations are measured in 
counter clockwise direction from the radial outward direction.

As a final cross-check, inserting the measured values for 
$\bsigma$ and $\matrix{A}$ into Eq.~(\ref{eq:bareps2}) leads
to the measured values of $\bvarepsilon$ within a few percent
deviation. However, the orientation of the deformation
gradient is not well reproduced -- it seems very sensitive to
small fluctuations.

\section{Summary and Conclusion}

Discrete element simulations of a 2D Couette shear cell were presented
and used as the basis for a micro-macro averaging procedure. 
In the shear cell a shear band is localized close to the inner,
rotating cylindrical wall.
The boundary conditions were chosen to allow for averaging over 
large volumes (rings with width $\Delta r$) and over a steady
state and thus over long times. The configurations changed rather
rapidly in the shear band, whereas the system is frozen in the
outer part, a fact which requires either extremely long simulations
or a sampling over different initial conditions in order to
allow an ensemble averaging in the outer part.  

The simplest averaging strategy involves only the par\-ticle-centers
as carriers of the quantities to be averaged over, whereas a more 
advanced method assumes the quantities to be homogeneously smeared
out over the whole particle which is cut in slices by the averaging
volumes. Both methods agree if the averaging volumes are of the 
particle size (or multiples), but for other sizes differing results 
are obtained. The slicing method shows discretization effects 
in the range of averaging volume widths from one to one fifth of 
a particle diameter, while the particle-center method shows 
fluctuations due to the choice of the averaging volume in 
a much wider range. 

The material density, i.e.~the volume fraction, the coordination 
number, the fabric tensor, the stress tensor and the elastic, 
reversible deformation gradient were obtained by the averaging
procedures.  The fabric is linearly proportional
to the product of volume fraction and coordination number.
In the shear band, dilation together with a reduction of
the number of contacts is observed. The mean stress is constant
in radial direction while the deformation gradient decays with
the distance from the inner wall. The ratio of the volumetric parts 
of stress and strain gives the effective stiffness of the granulate,
which is small in the shear band and larger outside, due to 
dilation. 

In the shear band, large deviatoric components of all tensorial 
quantities are found, however, decreasing with increasing
distance from the inner wall. The isotropy of the tensors grows
only slightly with increasing density and all tensors
are tilted counterclockwise from the radial direction by an angle 
of the order of $\pi/4$.
The system organizes itself such that more contacts are created
to act against the shear.
An essential observation is that the macroscopic tensors are {\em not}
co-linear, i.e.~their orientations are different. The orientation 
of the fabric is tilted most, that of the deformation gradient is 
tilted least and thus, the material cannot be described by a simple 
elastic model involving only two Lam\'e constants (or bulk modulus and 
Poisson's ratio) as the only parameters.  Alternatives are to cut
the system into pieces with different material properties and thus 
introducing discontinuities \cite{herrmann98}  or to use the rank four
stiffness tensor for anisotropic materials \cite{luding99c}.
The deviatoric parts of stress and deformation
gradient are seemingly interconnected via the fabric tensor.

To conclude, we proposed a consistent averaging formalism 
to obtain a mean quantity $Q$ in average over arbitrary volumes $V$. 
Within this framework, we used for $Q^c$
%the volume fraction  $\nu^c        =1/{\cal C}^p$,
the fabric           $\matrix{F}^c =\vec{n}^{c} \otimes \vec{n}^{c}$,
the stress           $\bsigma^c    =(1/V^p)\vec{f}^c \otimes \vec{l}^{pc}$, and
the deformation gradient $\bvarepsilon^c 
                                   =(\delta^c/a_p) \vec{n}^{c} \otimes 
                                      \vec{n}^{c} \cdot \matrix{A}$.
Future work will involve the extension of the present analysis 
to measure also the fluctuations of the above quantities and, in
addition, other interesting quantities like, e.g.~, the plastic strain
and non-symmetric parts of the stress tensor due to the effective
moments acting on single particles. More systematic parameter studies
are currently in progress.

%\bibliography{/home/lui/LIT/granulates,./granma}
%\bibliographystyle{unsrt}

\end{document}